\global\def\draftcontrol{0}  
   \def\versionno{ Regge Trajectories in String Duals of Confining Theories   }  
 
\catcode`\@=11 
 
\expandafter\ifx\csname draftcontrol\endcsname\relax\global\def\draftcontrol{0} 
\fi 
 
{\count255=\time\divide\count255 by 60 
\xdef\hourmin{\number\count255} 
\multiply\count255 by-60\advance\count255 by\time 
\xdef\hourmin{\hourmin:\ifnum\count255<10 0\fi\the\count255}} 
\def\draftdate{\number\month/\number\day/\number\year\ \ \ \hourmin } 
 
\newcommand\makepapertitle{\par 
  \begingroup 
    \renewcommand\thefootnote{\@fnsymbol\c@footnote}%
    \def\@makefnmark{\rlap{\@textsuperscript{\normalfont\@thefnmark}}}%
    \long\def\@makefntext##1{\parindent 1em\noindent 
            \hb@xt@1.8em{%
                \hss\@textsuperscript{\normalfont\@thefnmark}}##1}%
     \newpage 
     \global\@topnum\z@   
     \@makepapertitle 
     \thispagestyle{empty}\@thanks 
  \endgroup 
  \setcounter{footnote}{0}%
  \global\let\thanks\relax 
  \global\let\makepapertitle\relax 
  \global\let\@makepapertitle\relax 
  \global\let\@thanks\@empty 
  \global\let\@author\@empty 
  \global\let\@date\@empty 
  \global\let\@title\@empty 
  \global\let\title\relax 
  \global\let\author\relax 
  \global\let\date\relax 
  \global\let\and\relax 
  \def\version{\let\version\@version\@gobble} 
} 
\def\@makepapertitle{%
  \newpage 
   \ifnum\draftcontrol=1 {} 
   \version\versionno 
   \vskip 3em%
   \else 
   \hfill\hbox to 3cm {\parbox{4cm}{\@pubnum}\hss}%
   \vskip 3em%
   \fi 
   \begin{center}%
   \let \footnote \thanks 
     {\LARGE {\@title}}%
     \vskip 1.5em%
     {\normalsize
       \lineskip .5em%
       \begin{tabular}[t]{c}%
         \@author 
       \end{tabular}\par}%
     \vskip 1.5em%
     {\@bstract}%
     \end{center}%
     \vskip 1.5em  
     \@date%
   \par 
} 
 
\gdef\@pubnum{} 
\def\pubnum#1{%
  \gdef\@pubnum{#1}} 
 
\gdef\@bstract{} 
\def\Abstract#1{%
  \gdef\@bstract{%
   \parbox{\textwidth-0pc}{%
   \centerline{\bf Abstract}\penalty1000%
\noindent
\renewcommand\baselinestretch{1.0}%
{#1}}} 
}

\def\ps@paper{\let\@mkboth\@gobbletwo%
     \ifnum\draftcontrol=1 
        \def\@oddfoot{\hbox to \textwidth{\tiny \versionno \hfil\tiny\draftdate}%
        \hskip -\textwidth \hbox to \textwidth{\hfil\rm\thepage\hfil}}%
     \else\def\@oddfoot{\hbox to \textwidth{\hfil\rm\thepage\hfil}} 
     \fi 
     \let\@evenfoot\@oddfoot 
} 
 
\def\@version#1{\ifnum\draftcontrol=1 
\typeout{}\typeout{#1}\typeout{} 
\vskip3mm\centerline{\hbox{\fbox{\normalsize{\tt DRAFT -- #1 -- } 
                   {\draftdate}}}}\vskip3mm 
\fi} 
\let\version\@version 
\long\def\eqlabel#1{\ifnum\draftcontrol=1 
                    \tag@false  
                    \tag*{(\theequation) \hbox to -0.2cm{\hspace{0cm}\small{#1}\hss}} 
                    \refstepcounter{equation}  
                    \edef\@currentlabel{\theequation} 
                    \ltx@label{#1}          
                    \else 
                    \label{#1} 
                    \fi 
                    } 
\let\st@bibitem\@bibitem 
\let\st@lbibitem\@lbibitem 
\ifnum\draftcontrol=1 
  \def\@bibitem#1{%
    \st@bibitem{#1}\a@@label{#1}\ignorespaces} 
  \def\@lbibitem[#1]#2{%
    \st@lbibitem[#1]{#2}\a@@label{#2}\ignorespaces} 
  \def\a@@label#1{%
    \gdef\a@lab{\smash{\normalfont\small#1}} 
    \ifvmode 
      \if@inlabel 
        \global\setbox\@labels\hbox{%
          \llap{\a@lab\let\a@lab\relax 
                \kern\@totalleftmargin\kern\marginparsep}%
          \box\@labels}%
      \fi 
    \fi} 
\fi 

\documentclass[12pt,letterpaper]{article} 
 
\usepackage{amsmath,amssymb,array,calc,rotating,epsfig,psfrag}  
\usepackage[nosort]{cite}  
\ifnum\draftcontrol=1  
\fi  
\renewcommand\baselinestretch{1.25}  
\renewcommand\section{\@startsection {section}{1}{\z@}%
                             {-3.5ex \@plus -1ex \@minus -.2ex}%
                             {2.3ex \@plus.2ex}%
                             {\normalfont\large\bfseries}}  
\renewcommand\subsection{\@startsection{subsection}{2}{\z@}%
                             {-3.25ex\@plus -1ex \@minus -.2ex}%
                             {1.5ex \@plus .2ex}%
                             {\normalfont\normalsize\bfseries}}  
\renewcommand\subsubsection{\@startsection{subsubsection}{3}{\z@}%
                             {-3.25ex\@plus -1ex \@minus -.2ex}%
                             {1.5ex \@plus .2ex}%
                             {\normalfont\normalsize\it}}  
\renewcommand\paragraph{\@startsection{paragraph}{4}{\z@}%
                             {-3.25ex\@plus -1ex \@minus -.2ex}%
                             {1.5ex \@plus .2ex}%
                             {\normalfont\normalsize\bf}}   
   \def\revise#1       {\raisebox{-0em}{\rule{3pt}{1em}}%
               \marginpar{\raisebox{.5em}{\vrule width3pt\ 
               \vrule width0pt height 0pt depth0.5em 
               \hbox to 0cm{\hspace{0cm}{%
               \parbox[t]{4em}{\raggedright\footnotesize{#1}}}\hss}}}}

\def\calm         {{\cal M}}

   \def\del          {\partial}  
  
\def\ee           {{\rm e}}

\def\tr           {\mathop{\rm Tr}}

\def\id           {{\it id}}   \def\de#1#2{{\rm d}^{#1}\!#2\,}

\def\sqr#1#2{{\vcenter{\vbox{\hrule height.#2pt   
 \hbox{\vrule width.#2pt height#1pt \kern#1pt 
 \vrule width.#2pt}\hrule height.#2pt}}}}  

\def\a{\alpha}  
\def\b{\beta}  
\def\r{\rho}

\def\la{\lambda}   
\def\be{\begin{equation}}  
\def\ee{\end{equation}}   \def\m{\mu}  
\def\g{\gamma}  
\def\l{\lambda}  
\def\n{\nu}  
\catcode`\@=12   
\begin{document}   
     
\def \el {{\ell}}     
\def \KK {{\cal  K}}     
\def \K {{\rm K}}     
\def \tz{\tilde{z}}    
     
\def \ci {\cite}     
\newcommand{\rf}[1]{(\ref{#1})}     
\def \la {\label}     
\def \const {{\rm const}}

\def \ov {\over}     
\def \ha {\textstyle { 1\ov 2}}     
\def \we { \wedge}     
\def \P { \Phi} \def\ep {\epsilon}     
\def \ab {{A^2 \ov B^2}}     
\def \ba {{B^2 \ov A^2}}     
\def \tv   {{1 \ov 12}}     
\def \go { g_1}\def \gd { g_2}\def \gt { g_3}     
\def \gc { g_4}\def \gp { g_5}\def \F {{\cal F}}     
\def \del { \partial}     
\def \t {\theta}     
\def \p {\phi}     
\def \ep {\epsilon}     
\def \te {\tilde \epsilon}     
\def \ps {\psi}     
\def \x {{x_{11}}}        
\def\br{\bar{\rho}}     
\newcounter{subequation}[equation]

\def\pa{\partial}     
\def\e{\epsilon}     
\def\rt{\rightarrow}     
\def\tr{{\tilde\rho}}     
\newcommand{\eel}[1]{\label{#1}\end{equation}}     
\newcommand{\bea}{\begin{eqnarray}}     
\newcommand{\eea}{\end{eqnarray}}     
\newcommand{\eeal}[1]{\label{#1}\end{eqnarray}}     
\newcommand{\LL}{e^{2\lambda(r)}}     
\newcommand{\NN}{e^{2\nu(r)}}     
\newcommand{\PP}{e^{-2\phi(r)}}     
\newcommand{\non}{\nonumber \\}     
\newcommand{\CR}{\non\cr}

\makeatletter     
     
\def\thesubequation{\theequation\@alph\c@subequation}     
\def\@subeqnnum{{\rm (\thesubequation)}}     
\def\slabel#1{\@bsphack\if@filesw {\let\thepage\relax    
   \xdef\@gtempa{\write\@auxout{\string    
      \newlabel{#1}{{\thesubequation}{\thepage}}}}}\@gtempa    
   \if@nobreak \ifvmode\nobreak\fi\fi\fi\@esphack}     
\def\subeqnarray{\stepcounter{equation}     
\let\@currentlabel=\theequation\global\c@subequation\@ne     
\global\@eqnswtrue \global\@eqcnt\z@\tabskip\@centering\let\\=\@subeqncr    
     
$$\halign to \displaywidth\bgroup\@eqnsel\hskip\@centering    
  $\displaystyle\tabskip\z@{##}$&\global\@eqcnt\@ne    
  \hskip 2\arraycolsep \hfil${##}$\hfil    
  &\global\@eqcnt\tw@ \hskip 2\arraycolsep    
  $\displaystyle\tabskip\z@{##}$\hfil    
   \tabskip\@centering&\llap{##}\tabskip\z@\cr}     
\def\endsubeqnarray{\@@subeqncr\egroup    
               $$\global\@ignoretrue}     
\def\@subeqncr{{\ifnum0=`}\fi\@ifstar{\global\@eqpen\@M    
    \@ysubeqncr}{\global\@eqpen\interdisplaylinepenalty \@ysubeqncr}}     
\def\@ysubeqncr{\@ifnextchar [{\@xsubeqncr}{\@xsubeqncr[\z@]}}     
\def\@xsubeqncr[#1]{\ifnum0=`{\fi}\@@subeqncr    
   \noalign{\penalty\@eqpen\vskip\jot\vskip #1\relax}}     
\def\@@subeqncr{\let\@tempa\relax    
    \ifcase\@eqcnt \def\@tempa{& & &}\or \def\@tempa{& &}    
      \else \def\@tempa{&}\fi    
     \@tempa \if@eqnsw\@subeqnnum\refstepcounter{subequation}\fi    
     \global\@eqnswtrue\global\@eqcnt\z@\cr}     
\let\@ssubeqncr=\@subeqncr     
\@namedef{subeqnarray*}{\def\@subeqncr{\nonumber\@ssubeqncr}\subeqnarray}    
     
\@namedef{endsubeqnarray*}{\global\advance\c@equation\m@ne    
                     \nonumber\endsubeqnarray}    
     
\makeatletter \@addtoreset{equation}{section} \makeatother     
\renewcommand{\theequation}{\thesection.\arabic{equation}}     
     
\def \ci {\cite}     
\def \la {\label}     
\def \const {{\rm const}}     
\catcode`\@=11    
     
\newcount\hour     
\newcount\minute     
\newtoks\amorpm \hour=\time\divide\hour by 60\minute     
=\time{\multiply\hour by 60 \global\advance\minute by-\hour}     
\edef\standardtime{{\ifnum\hour<12 \global\amorpm={am}    
  \else\global\amorpm={pm}\advance\hour by-12 \fi    
  \ifnum\hour=0 \hour=12 \fi    
  \number\hour:\ifnum\minute<10    
  0\fi\number\minute\the\amorpm}}     
\edef\militarytime{\number\hour:\ifnum\minute<10 0\fi\number\minute}    
     
\def\draftlabel#1{{\@bsphack\if@filesw {\let\thepage\relax    
   \xdef\@gtempa{\write\@auxout{\string    
      \newlabel{#1}{{\@currentlabel}{\thepage}}}}}\@gtempa    
   \if@nobreak \ifvmode\nobreak\fi\fi\fi\@esphack}    
  \gdef\@eqnlabel{#1}}     
\def\@eqnlabel{}     
\def\@vacuum{}     
\def\marginnote#1{}     
\def\draftmarginnote#1{\marginpar{\raggedright\scriptsize\tt#1}}     
\overfullrule=0pt    
    
 \def \lc {light-cone\ }    
     
\def\draft{    
  \pagestyle{plain}    
  \overfullrule=2pt    
  \oddsidemargin -.5truein    
  \def\@oddhead{\sl \phantom{\today\quad\militarytime} \hfil    
  \smash{\Large\sl DRAFT} \hfil \today\quad\militarytime}    
  \let\@evenhead\@oddhead    
  \let\label=\draftlabel    
  \let\marginnote=\draftmarginnote    
  \def\ps@empty{\let\@mkboth\@gobbletwo    
  \def\@oddfoot{\hfil \smash{\Large\sl DRAFT} \hfil}    
  \let\@evenfoot\@oddhead}    
     
\def\@eqnnum{(\theequation)\rlap{\kern\marginparsep\tt\@eqnlabel}    
  \global\let\@eqnlabel\@vacuum}  }    
     
\renewcommand{\rf}[1]{(\ref{#1})}     
\renewcommand{\theequation}{\thesection.\arabic{equation}}     
\renewcommand{\thefootnote}{\fnsymbol{footnote}}    
     
\newcommand{\newsection}{    
\setcounter{equation}{0}     
\section}    
     
\textheight = 22truecm      
\textwidth = 17truecm      
\hoffset = -1.3truecm      
\voffset =-.5truecm    
     
\def \tx {\textstyle}     
\def \tix{\tilde{x}}     
\def \bi{\bibitem}    
     
\def \ov {\over}     
\def \ha {\textstyle { 1\ov 2}}     
\def \we { \wedge}     
\def \P { \Phi} \def\ep {\epsilon}     
\def \ab {{A^2 \ov B^2}}     
\def \ba {{B^2 \ov A^2}}     
\def \tv   {{1 \ov 12}}     
\def \go { g_1}\def \gd { g_2}\def \gt { g_3}     
\def \gc { g_4}\def \gp {     
g_5}     
\def \F {{\cal F}}     
\def \del { \partial}     
\def \t {\theta}     
\def \p {\phi}     
\def \ep {\epsilon}     
\def \ps {\psi}

\def \LL{{\cal L}}     
\def\o{\omega}     
\def\O{\Omega}     
\def\e{\epsilon}     
\def\pd{\partial}     
\def\pdz{\partial_{\bar{z}}}     
\def\bz{\bar{z}}     
\def\e{\epsilon}     
\def\m{\mu}     
\def\n{\nu}     
\def\a{\alpha}     
\def\b{\beta}     
\def\g{\gamma}     
\def\G{\Gamma}     
\def\d{\delta}     
\def\r{\rho}     
\def\bx{\bar{x}}     
\def\by{\bar{y}}     
\def\bm{\bar{m}}     
\def\bn{\bar{n}}     
\def\s{\sigma}     
\def\na{\nabla}     
\def\D{\Delta}     
\def\l{\lambda}     
\def\te{\theta} \def \t {\theta}     
\def\ta {\tau}     
\def\na{\bigtriangledown}     
\def\p{\phi}     
\def\L{\Lambda}     
\def\hR{\hat R}     
\def\ch{{\cal H}}     
\def\ep{\epsilon}     
\def\bj{\bar{J}}     
\def \foot{ \footnote}     
\def\be{\begin{equation}}     
\def\ee{\end{equation}}     
\def \P {\Phi}     
\def\un{\underline{n}}     
\def\ur{\underline{r}}     
\def\um{\underline{m}}     
\def \ci {\cite}     
\def \g {\gamma}     
\def \G {\Gamma}     
\def \k {\kappa}     
\def \l {\lambda}     
\def \L {{L}}     
\def \Tr {{\rm Tr}}     
\def\apr{{A'}}     
\def \m {\mu}     
\def \n {\nu}     
\def \W{{\cal W}}     
\def \eps {\epsilon}     
\def \ha{{     
 { 1 \ov 2}} }     
\def \de{{     
{ 1 \ov 9}} }     
\def \si{{     
 { 1 \ov 6}} }     
\def \fo{{     
{ 1 \ov 4}} }     
\def \ei{{     
{ 1 \ov 8}} }     
\def \rt {{\tx { \ta \ov 2}}}     
\def \rr {{\bar \rho}}    
     
\def\D{\Delta}     
\def\l{\lambda}     
\def\L{\Lambda}     
\def\te{\theta}     
\def\g{\gamma}     
\def\Te{\Theta}     
\def\tw{\tilde{w}}

\def\sn{\rm sn}   
\def\cn{\rm cn}    
\def\dn{\rm dn}    
     

\topmargin=0.50in   
     
\date{}     
     
\begin{titlepage}    
     
\version\versionno   
     
\hfill hep-th/0311190    
     
\hfill MCTP-03-45  
  
\hfill PUPT-2100     
 
\begin{center}     
     
{\Large \bf Regge Trajectories Revisited }       
\vskip .4cm    
{\Large \bf in the Gauge/String Correspondence}          
\vskip .6 cm    
     
{\large   Leopoldo A. Pando Zayas${}^{1}$, Jacob Sonnenschein${}^{2}$    
and Diana Vaman${}^3$ }\\    
     
\end{center}    
     
\vskip .2cm  
\centerline{\it ${}^1$ Michigan Center for Theoretical     
Physics}     
\centerline{ \it Randall Laboratory of Physics, The University of     
Michigan}     
\centerline{\it Ann Arbor, MI 48109-1120}     
     
\vskip .2cm  
\centerline{\it ${}^2$ School of Physics and Astronomy}     
\centerline{ \it Beverly and Raymond Sackler Faculty of Exact Sciences}     
\centerline{ \it Tel Aviv University, Ramat Aviv, 69978, Israel}    
     
\vskip .2cm \centerline{\it ${}^3$ Department of Physics}    
\centerline{ \it Princeton University}    
\centerline{\it Princeton, NJ 08544}


\begin{abstract}     
We attempt to obtain realistic glueball Regge trajectories from the 
gauge/string correspondence. To this end we study closed spinning string 
configurations in two supergravity backgrounds: Klebanov-Strassler (KS) and 
Maldacena-Nunez (MN) which are dual to confining gauge theories.  
These backgrounds represent two embeddings of 
${\cal N}=1$ SYM, in the large rank $N$ limit, in string theory.  
The classical configuration that 
we consider is that of a folded  closed string spinning in a  
supergravity region with vanishing transverse radius ($\tau = 0$)  
which is dual to the IR of the gauge theory.  
Classically, a spinning  string yields a linear Regge 
trajectory with zero intercept. By performing a semi-classical analysis we 
find that quantum effects alter both the linearity of the trajectory and 
the vanishing classical intercept: $J\equiv \alpha(t)= 
\alpha_0 + \alpha' t +\beta\sqrt{t}$. Two features of our Regge trajectories are compatible with the  
experimental  Pomeron trajectory: positive intercept and positive 
curvature. The fact that both KS and MN string backgrounds give the 
same functional expression of the Regge trajectories suggests that in fact 
we are observing string states dual to ${\cal N}=1$ SYM. 
\end{abstract}

\end{titlepage}     
\setcounter{page}{1} \renewcommand{\thefootnote}{\arabic{footnote}}     
\setcounter{footnote}{0}    
     
\def \N{{\cal N}}      
\def \ov {\over}    
     
\tableofcontents    
 
\pagebreak

\section{Introduction}    
Regge theory is concerned  
with the particle spectrum, the forces between particles, and the high  
energy behavior of scattering amplitudes (for a comprehensive review  
see \cite{collins}).   
A first principle explanation  
of this theory remains an outstanding challenge for high energy particle  
theory. One of the most distinctive features of Regge theory are the  
Regge trajectories. A Regge trajectory is a line in a Chew-Frautschi \cite{cf} 
plot representing the spin of the lightest particles  
of that spin versus their mass$^2=t$:  
$\,\,J=\alpha_0+ \alpha'\,t$.

The fact that Regge trajectories are well described by simple string  
models has  
been known almost since the experimental verification of the Regge  
trajectories. The string models used to describe Regge trajectories are  
generically related to strings in flat space and beyond this particular  
feature they do not provide a framework for describing hadronic  
physics.  Moreover, there is no fundamental reason for a string in flat  
space to be useful in the  
description of some properties of hadronic states in a gauge theory.

Not long ago a precise duality between gauge theories and string theories  
has been uncovered \cite{malda}. In particular,   
in the context of the AdS/CFT correspondence backgrounds have been  
constructed that represent supergravity duals of confining gauge  
theories. In some cases the precise field theoretic content of the  
theory is known  \cite{ks,mn}. The Klebanov-Strassler (KS)  
supergravity  solution,   
for example, contains in a geometric  
way information about confinement, chiral symmetry breaking,  
duality cascade and   instanton effects among others. Having a string  
theory that mathematically contains the same information  
as a confining gauge  theory provides us, in principle, with the means  
to study the IR dynamics of confining gauge theories. Indeed, some of the  
hadronic states of confining gauge theories which admit a supergravity  
dual description have recently been described \cite{gpss}. In this paper  
we use the gauge/gravity framework to revisit one of the trademark  
properties of confined matter -- Regge trajectories.

A central question in our analysis is what are the precise features of  
the new description of confining theories that differ from the more  
naive approach of describing Regge trajectories via strings in flat  
space. We will show that at the classical level strings in confining  
backgrounds behave exactly as strings in flat space. It is only at  
one-loop level where the difference becomes significant. In  
particular, we will show that the intercept of Regge trajectories is a  
one-loop effect and therefore differs for superstrings in flat space and  
strings in confining backgrounds.

The traditional approach to obtaining Regge trajectories from string  
theories relies on identifying the spectrum of  classical   
strings spinning in flat space with the spectrum of the corresponding  
particles.   
Recently, however, in the  
context of the gauge/gravity correspondence the role of classical  
solutions has been reexamined. It has been proposed that  
classical solutions of the string sigma model are in correspondence with   
sectors of large quantum numbers in the dual gauge theory \cite{gkp}.   
As a concrete example,   
\cite{gkp} considered a closed string spinning   
in AdS and established its correspondence with twist-two operators in  
${\cal N}=4$ SYM. Namely, they obtained that the conserved quantities of  
the solitonic solution in the sigma model when translated in terms of  
gauge theory quantities imply that $\Delta -S =(\sqrt{\lambda}/\pi)\ln S$  
which is a prediction for the anomalous dimension of twist-two operators in ${\cal N}=4$  
SYM at large 't Hooft coupling. Note that  except for the dependence on the 't Hooft coupling  
$\lambda$, this is the form of the anomalous dimension for twist-two  
operators in the perturbative regime. A very powerful feature of the  
correspondence \cite{gkp} is that it allows a framework to go beyond the  
classical level. One can compute quantum corrections to the classical  
relation among the conserved quantities in the worldsheet approach and  
they should be in correspondence with quantum corrections on the gauge  
theory side. Indeed, Frolov and Tseytlin \cite{ft} performed the  
one-loop analysis of the classical solution proposed in \cite{gkp} and  
found that to this level terms of the form $\ln^2 S $ are absent,  
something that has been conjectured in the field theory approach to  
twist-two operators.

Under the point of view of \cite{gkp}, the large R-charge sector of  
${\cal N}=4$ SYM described in  \cite{bmn} can be seen to be dual on the  
string theory side to a classical   
string shrunk to a point an orbiting along the great circle of  $S^5$  
in the $AdS_5\times S^5$ background. In this case  the  
description of \cite{bmn} has the advantage, over the description of   
\cite{gkp}, of being exact on the string  
theory side.  Using a limit similar to \cite{bmn},  an exact description  
of a set of hadrons with large flavor charged (annulons) was found in \cite{gpss}   
(similar states were also found in nonsupersymmetric  
theories \cite{cotrone,stanislav}).  
This sector can correspondingly be described  
as a classical configuration describing an extended string located near  
values of the radial direction in the Sugra background that correspond  
to the IR region of the gauge theory. This configuration also carries   
large angular momentum in the internal space perpendicular to the world  
volume. The density of states for these hadrons was 
computed in \cite{hag} and found to be of Hagedorn type with a 
coefficient determined by the quark-antiquark string tension and their 
flavor charge.  
The annulons  are typical of   
embeddings of ${\cal N}=1$  
SYM into string theory but their quantum numbers are not shared by more  
realistic theories like QCD.  
    
With this improved understanding of the role of classical solutions in  
the gauge/gravity correspondence and with the hope of describing states  
in the same universality class as the hadrons of QCD, we turn to the study of   
strings spinning in supergravity backgrounds dual to confining gauge  
theories. Our main motivation is to obtain a description of Regge  
trajectories in a situation where the relation between string theory  
and gauge theory has been established fairly rigorously. We study the  
classical configurations in general confining backgrounds and then  
proceed to explicitly consider the semiclassical quantization in  the  
case of the KS and MN backgrounds.

Our analysis is semiclassical in nature, that is we compute {\it quantum}   
corrections to a given {\it classical} relation. This situation  
differs from the description of the annulons \cite{gpss} in which by  
using an appropriate Penrose limit one obtains an  
exact string theory. Nevertheless, the robustness of our semiclassical  
approach is well justified. In fact, there is an extensive list of  
examples where a similar semiclassical approach has been used yielding  
very reliable results \cite{ft,t,ftn,tinteg}.    
 
This paper is organized as follows.  
In section \ref{classical} we review some of the standard classical  
solutions that are relevant for the study of Regge trajectories,  
including open and closed strings spinning in flat space. We also  
consider closed strings in confining backgrounds (some classical 
aspects were considered in \cite{pepe}). We begin by solving 
the classical equation of motion for strings in generic confining 
supergravity backgrounds: our string configuration describes a  
string spinning in the region dual to the IR gauge theory. 
 
Section \ref{q}  
contains an account of the quantization of strings spinning in flat  
space. In the phenomenological literature several effective string  
models have been proposed as possible sources of corrections to Regge  
trajectories. To our knowledge, we present the first such analysis in  
the context of IIB string theory. We find that in flat space the string  
spinning is a BPS configuration and therefore receives no corrections at  
the one loop level. For the Regge trajectories this implies that the  
intercept is that of the classical trajectories, that is 
zero. However, considering the bosonic contribution alone  one finds a 
linear trajectory with positive intercept, something in qualitative 
agreement with experimental data.   
 
In section \ref{quadra},  we  
consider quadratic fluctuations  
around a classical  string spinning in the two trademark  
supergravity backgrounds dual  
to confining theories: KS and MN.  
Interestingly,  we find that from the perspective of a semiclassical 
quantization around the spinning string configuration, both theories behave  
exactly the same way. We therefore present in section \ref{qcrt} a  
unified analysis of the quantum corrected Regge trajectories.   
 
Section \ref{pheno} contains our  
attempts to compare our results with phenomenological data about Regge  
trajectories available from direct experiments or lattice  
calculations. In particular, we emphasize the relevance of our  
calculations for the soft Pomeron trajectory and the lattice results for  
glueballs. It is worth mentioning that although the supergravity  
theories that we considered are dual to embeddings of ${\cal N}=1$ SYM  
into IIB string theory, the comparison with results of QCD shows that for  
questions pertaining to Regge trajectories these theories are certainly  
in the same universality class.        
 
\section{Classical Regge trajectories}  
\label{classical} 
   
Regge trajectories are very generic in hadronic physics. For example,  
they are very well established for mesons, baryons and the soft Pomeron  
\cite{collins} (see also section \ref{pheno}). It is remarkable that the  
gauge/gravity correspondence provides a framework in which each of  
these states can be shown to be described by a specific classical  
solution.   
\begin{table}[hbt]  
\begin{center}      
\begin{tabular}[h]{|l||l|}      
\hline     
 Gauge Theory State & String Theory Configuration \\     
\hline       
\hline       
Glueballs  & Spinning Folded Closed String \\      
\hline       
Mesons of heavy quarks & Spinning open strings \\      
\hline       
Baryons of heavy quarks  &  Strings attached to a baryonic vertex \\      
\hline       
Dibaryons  &  Strings attached to wrapped branes \\      
\hline       
\end{tabular}      
\caption{States in gauge theory and their corresponding classical 
 configuration  in the string theory.}  
\end{center}      
\end{table}  
 
An important point in the correspondence is the need to  consider objects with  
angular momentum in Poincare coordinates. This implies that the  
conserved quantity conjugate  to the Poincare time ($E$) 
measures the energy of a  
state in Minkowski space, that is, one is   
considering the energy of a state in the gauge theory side rather than the conformal dimension  
as corresponds to a configuration in global AdS time.   
The main feature  of these classical solutions is that  
they are characterized (among other classical quantities) by their world  
volume angular  
momentum ($J$) which is the gravity dual to spin in the gauge theory.     
 
There is a distinctive feature that distinguishes between  
configurations of closed and open strings with angular momentum. Open  
strings with angular momentum are characterized by $E$,  $J$ and $L$ 
(with $L$ the endpoints separation),  
whereas closed string with angular momentum are only characterized by  
$E$ and $J$.         
\subsection{Spinning open string in flat space}   
 
Let us briefly review the familiar story of the Regge trajectories  
associated with the spinning string in flat space. We begin with   
the Polyakov action restricted to the bosonic degrees of freedom of the string 
\be 
S=\frac{T_s}2 \int_0^\pi d\sigma \int d\tau \sqrt{\g} \g^{\a\b} 
\partial_\a X^\mu 
\partial_\b X_\mu =\frac{T_s}2 \int_0^\pi d\sigma \int d\tau L(\s,\tau) 
\label{action1}  
\ee 
where the string tension is $T_s=1/(2\pi\a')$. 
The spinning string classical configuration is described by 
\be 
X^0 = e\tau , \qquad X^1 = e\cos\s\cos\tau , \qquad 
X^2 = e\cos\s\sin\tau 
\ee 
This solution satisfies the equations of motion of the coordinates 
$X^\mu(\s,\tau)$, the Virasoro constraints, and the Neumann boundary 
conditions $\partial_\s X^\mu |_{\s=0,\pi}=0$. 
 
The energy and angular momentum associated with this classical configuration 
are defined as 
\bea 
E&=&T_s\int_0^\pi d\s \frac{\partial X^0}{\partial \tau}=\pi T_s e\nonumber\\ 
J&=& \int_0^\pi d\s (\dot {X}^2 X^1- \dot X^1 X^2)=  \frac {\pi}2 T_s e^2 
\eea 
The Regge trajectory is defined as the relationship $J=J(E^2)$ 
\be 
\eqlabel{rego} 
J=\frac{1}{2\pi T_s} E^2 = \a' E^2 \equiv \alpha' t 
\ee 
The slope of the Regge trajectory of a spinning open string is therefore 
$\a'$. Note that from the standpoint of a classical configuration the  
fermionic degrees are naturally set to zero and they do not affect the  
form of the Regge trajectory.

\subsection{Spinning closed  string in flat space}   
 
For a closed string, we have to require periodicity in $\s$ on the  
interval $[0,2\pi]$. Therefore the classical spinning closed string  
configuration, solution to the equations of motion derived from the action 
(\ref{action1}), will be 
\be 
X^0 =e\tau , \qquad X^1 =  e \sin(\s)\cos(\tau) , \qquad 
X^2 =  e\sin(\s)\sin(\tau)\label{closedclass} 
\ee 
Note that the solution (\ref{closedclass}) obeys the Virasoro constraints as  
well. We chose the particular dependence on $\sigma$ such that the center 
of the string be at rest. 
 
The energy and angular momentum of this configuration are 
\bea 
E&=& T_s\int_0^{2\pi} d\s \frac{\partial X^0}{\partial \tau}=2\pi T_s  
e\nonumber\\ 
J&=& T_s\int_0^{2\pi} d\s (\dot {X}^2 X^1- \dot X^1 X^2)=  \pi T_s {e^2} 
\eea 
 
The Regge trajectory of a closed string 
\be 
\eqlabel{regc} 
J=\frac{1}{4\pi T_s} E^2 = \frac{\a'}2 E^2 \equiv \frac 12 \alpha' t 
\ee 
will have a slope equal to a half  the Regge slope of an open string.  
The difference between the slope of Regge trajectories for mesons and  
glueball has been noticed phenomenologically. In section (\ref{pheno})  
we comment on the string theory result predicting a ration of a half and  
the most common field theoretic approach based on effective  
descriptions.      
\subsection{Spinning Wilson line  in flat space time}    
 
Some of the theories we consider do not admit ``open strings'' as  
microscopic degrees of freedom. However, they do admit open strings  
stretched between two fixed points in target space. In the context of   
the holographic duality  
these string configurations are the duals of the  Wilson loops of the  
corresponding gauge theory\cite{wl}.  Let us then consider such strings  
in flat space and in more general supergravity backgrounds dual to confining  
theories.    Since we are considering classical aspects only we choose to work with  
the Nambu-Goto action describing a string in flat space   
\be    
S_{NG}= \frac{1}{ 2\pi\alpha'}\int d^2\s \sqrt{-det(\g_{\alpha\beta})}   
=  \frac{1}{2\pi\alpha'}\int d^2\s   
\sqrt{\big[(\dot X^0)^2- \rho^2(\dot \phi)^2\big] (\rho')^2  },     
\ee     
where $\g_{\alpha\beta}$ is the induced worldsheet metric.   
For convenience we choose to  
work in polar coordinates which are related to the Cartesian coordinates  
by $X^1= \rho \cos\phi$ and $X^2= \rho \sin\phi$.    
A static Wilson loop is described by the configuration    
$X^0=e\tau,\,\, \rho= \frac{L}{\pi}\s,\,\,  \p=\p_0  $     
The space-time energy associated with this configuration  
\be    
E_{stat}= \frac{ 1}{2\pi\alpha'} \int d\sigma   
\frac{\partial X^0}{\partial\tau} =\frac{1}{2\pi\a'} L,    
\ee    
since by the Virasoro constraint $e= \frac{L}{\pi}$.   
This  linear potential obviously resembles a confining potential.     
For a spinning Wilson line we take the 
  ansatz $X^0=e\tau,\,\, \t=\pi/2,\,\, \p=e\o\tau, \,\, \rho=\rho(\s)$.  
The space-time    energy  is given by    
\be    
E=\frac{2}{ 2\pi\alpha'}\frac{1}{ \o} \arcsin ( L\o),    
\ee    
and the spin is     
\be    
J=\frac {2}{2\pi\alpha'} \int d\sigma \rho^2\frac{\p}{\tau} =  \frac{1}{2\pi\alpha'}  
\frac{1}{ \o^2}\big[ \arcsin  L\o-( L\o)\sqrt{1-( L\o)^2}\big].    
\ee    
The two in the numerator has the same origin as in the previous subsection.   
For the special case of $wL=1$ (when the ends of the strings move at  
the speed of light) we get    
\be    
E= \frac{1}{ 2\alpha'}\frac{1}{ \o},\qquad J= \frac{1}{ 4\alpha'}\frac{1}{ \o^2}.   
\ee    
such that we get the Regge behavior    
\be    
J = \a' E^2 \equiv \alpha' t 
\ee    
In the regime where  $\o L\sim 0 $ we get a correction to the linear    
term of the form    
\be    
\eqlabel{wlzero}  
E\approx  \frac{1}{ \pi\a'}\left( L +{3\pi^2 \over 2}\frac{\a'^2 J^2}{ L^3}\right).    
\ee      
Notice the positive sign of the second term that indicates that the  
rotation of the Wilson loop increases the binding energy of the quark  
anti-quark system. It is important to bear in mind that the sub-leading  
term in expression (\ref{wlzero}) represents classical corrections to the  
standard confining result. In addition there are quantum corrections  
that will be discussed in section  4. One interesting feature of the  
above relations is the possibility to interpolate between a linear  
potential between a quark and an antiquark and for small $\o L$ and Regge  
trajectory for $\o L \to 1$.

\subsection{Closed spinning strings in supergravity backgrounds}  
 
Our starting point would be a supergravity solution of a  
form that naturally generalizes the  $AdS_5$ metric   
in Poincare coordinates. Tacitly we assume that we are working with IIB  
SUGRA backgrounds and that the metrics we consider are appropriate  
deformations of $AdS_5$. We consider background metric that preserves Poincare invariance  
in the coordinates $(X^0,X^i)$. We will not dwell at this point in the  
specifics of the space transverse to the world volume of the ``D3''  
brane and simply denote most of its structure by ellipsis in the  
background metric:      
\be  
ds^2=h(r)^{-1/2}\bigg[-(dX^0){}^2 + dX_1^2+ dX_2^2+ dX_3^2\bigg]   
+ h(r)^{1/2}dr^2+\ldots  
\ee  
The relevant classical equations of motion for the string sigma model in  
this background are   
\begin{eqnarray}  
\pd_a(h^{-1/2}\eta^{ab}\pd_b X^0)&=&0, \nonumber \\  
\pd_a(h^{-1/2}\eta^{ab}\pd_b X^i)&=&0, \nonumber \\  
\pd_a(h^{1/2}\eta^{ab}\pd_b r)&=&{1\over 2} \pd_r(h^{-1/2})  
\eta^{ab}\big[-\pd_a X^0\pd_b X^0 + \pd_a X_i \pd_b X^i\big]. \nonumber \\  
\end{eqnarray}  
They are supplemented by the standard Virasoro constraints.  
We will attempt to construct spinning strings by taking the   
following ansatz  
\begin{eqnarray}  
\label{ansatz}  
X^0&=&e\, \tau, \nonumber \\  
X^1&=&f_1(\tau)\,g_1(\sigma), \qquad X^2=f_2(\tau)\,g_2(\sigma), \nonumber \\  
X^3&=& \rm{constant}, \qquad r=r(\s).  
\end{eqnarray}  
With this ansatz the equation of motion for $X^0$ is trivially satisfied.   
Let us first show that the form of the functions $f_i$ is   
fairly generic. Namely, considering the equation of motion for $X^i$ we   
obtain  
\be  
-h^{-1/2}g_i\ddot f_i + f_i \pd_\s(h^{-1/2}g_i')=0,  
\ee  
where a dot denotes a derivative with respect to $\tau$ and   
a prime denotes a derivative with respect to $\s$.  
Enforcing a natural separation of variables we see that     
\be  
\ddot f_i+(e\,\o)^2 f_i=0, \qquad \pd_\s(h^{-1/2}g_i')+(e\,\o)^2 h^{-1/2}g_i=0.  
\ee  
The radial equation of motion is   
\be  
\pd_\s(h^{1/2}\pd_\s\, r)={1\over 2}\pd_r(h^{-1/2})\big[e^2  
-g_i^2\dot{f}_i^2 + f_i^2 g'{}_i^2\big].  
\ee  
Finally the constraint becomes  
\be  
h^{1/2}r'^2+h^{-1/2}\big[-e^2+g_i^2\dot{f}_i^2 + f_i^2 g'{}_i^2\big]=0.  
\ee  
The integrals of motion we would like to consider are  
\begin{equation}  
\eqlabel{e}  
E={e\over 2\pi \a'}\int h^{-1/2}  d\s,   
\end{equation}  
\be  
J={1\over 2\pi \a'} \int h^{-1/2}  
\big[x_1\pd_\tau x_2 -x_2\pd_\tau x_1\big] d\s   
={1\over 2\pi \a'} \int h^{-1/2}g_1g_2\big[f_1\pd_\tau f_2 -   
f_2\pd_\tau f_1 \big] d\s  
\ee   
The above system can be greatly simplified by further taking   
\be  
f_1=\cos e\o\,\tau, \quad f_2=\sin e\o\, \tau, \quad   
\rm{and} \quad g_1=g_2.  
\ee  
Under these assumptions the equation of motion for $r$ and the   
Virasoro constraint become  
\begin{equation}  
\eqlabel{eom}  
\begin{split}  
&\pd_\s\left(h^{1/2}\,\pd_\s\,r\right)-{1\over 2}\pd_r(h^{-1/2})  
\big[e^2-(e\o)^2\,g^2 +g'{}^2\big]=0,  \\  
&h^{1/2}r'{}^2+h^{-1/2}\big[-e^2+(e\o)^2\, g^2 + g'{}^2\big]=0.  
\end{split}  
\end{equation}  
The angular momentum is then   
\begin{equation}  
\eqlabel{j}  
J={e\o\over 2\pi \a'} \int h^{-1/2}g^2 d\s.  
\end{equation}  
According to the gauge/gravity correspondence and in particular to the  
insight put forward in \cite{gkp}, this spinning string describe a state  
in the dual gauge theory with the same quantum numbers. Since we are  
working in Poincare coordinates the quantity canonically conjugate to  
time is the energy of the corresponding state in the four dimensional  
theory. The angular momentum of the string describes the spin of the  
corresponding state. Thus a spinning string in the Poincare coordinates  
is dual to a state of energy $E$ and spin $J$. In order for our  
semiclassical approximation to be valid we need the value of the action  
to be large, this imply that we are considering gauge theory states in  
the IR region of the gauge theory with large spin and large energy.   
We will show that, in the cases we study, the expressions (\ref{e}) and (\ref{j}) yield a  
dispersion relation that can be identified with  Regge  
trajectories. Moreover, we will compute semiclassically quantum  
corrections to the Regge trajectories using the dual string theories.     
\subsubsection{Closed spinning strings  in confining backgrounds}    
\label{stringconfine} 
    
Let us show that there exists a simple solution of the equations of  
motion (\ref{eom}) for any gravity background dual to a confining  
gauge theory. Recall that the conditions for a SUGRA background to be  
dual to a confining theory have been exhaustively explored  
\cite{cobiwl}. The main idea is to translate the condition for the vev  
of the rectangular Wilson loop to display an area law into properties  
that the metric of the supergravity background must satisfy. It has  
been established that one of the sufficient conditions is for $g_{00}$  
to have a nonzero minimum at some point $r_0$ usually known as the   
end of the space wall \cite{cobiwl}.  Note that precisely these two  
conditions ensure the existence of  a solution of (\ref{eom}). Namely,  
since  
$g_{00}=h^{-1/2}$ we see that for a point $r=r_0={\rm constant}$ is a solution if   
\begin{equation}  
\eqlabel{confconds}  
\del_r(g_{00})|_{r=r_0}=0, \qquad g_{00}|_{r=r_0}\ne 0.  
\end{equation}  
The first condition solve the first equation in (\ref{eom}) and the  
second condition makes the second equation nontrivial. Interestingly  
the second condition can be interpreted as enforcing that the  
quark-antiquark string tension be nonvanishing. It is worth mentioning   
that due to the UV/IR correspondence  
in the gauge/gravity duality the radial direction is  identified with  
the energy scale. In particular, $r\approx r_0$ is the gravity  
dual of the IR in the gauge theory. Thus, the string we are considering  
spins in the region dual to the IR of the gauge theory. Therefore we can  
conclude that it  
is dual to states in the field theory that are characteristic of the  
IR.  
     Let us now explicitly display the Regge trajectories. The classical  
solution is given by (\ref{ansatz}) with  $g(\s)$ solving the second  
equation from (\ref{eom}), that is, 
 $g(\s)=(1/\o)\sin(e\o\s)$. Imposing the periodicity  
 $\s\rightarrow \s+ 2\pi$ implies that  $e\o=1$ and hence 
\be  
\eqlabel{clasreg}  
X^0=e\,\tau, \qquad   
X^1=e\cos \,\tau \, \sin \,\s,\, \qquad X^2=e\sin \,\tau \, \sin \,\s.  
\ee  
The expressions  
for the energy and angular momentum of the string states are:  
\begin{equation}  
E=4 { e\, g_{00}(r_0)\over 2\pi \a'}\int d\sigma=2\pi{g_{00}(r_0)} T_s e ,  
\qquad J=4{  
g_{00}(r_0) e^2\over 2\pi \a'}\int \sin^2\s  d\sigma= \pi{g_{00}(r_0)} T_s e^2 .  
\end{equation}  
 
Defining now ${T_{s,~eff}}=g_{00}(r_0)/(2\pi  
\a')$  and $\a'{}_{eff}=\a'/g_{00}$  
we find that the  Regge trajectories take the form  
\begin{equation}  
\eqlabel{Regge}  
J={1\over 4\pi {T_{s,~eff}}} E^2 = {1\over 2} \alpha'{}_{eff} \,\,t.  
\end{equation}  
Notice that the main difference with respect to the result in flat space  
is that the slope is modified to $\a'{}_{eff}=\a'/g_{00}$.   
It is expected that a confining background will have states that  
align themselves in Regge  trajectories. The main purpose of our  
investigation is not the relation (\ref{Regge}) itself but rather the  
corrections that it receives and that can be computed explicitly in  
the gauge/gravity correspondence for specific backgrounds.       Of course the conditions  
(\ref{confconds}) are necessary conditions for confinement but they are  
not sufficient. Namely, there are backgrounds satisfying  
(\ref{confconds}) that are not dual to confining gauge theories.  The most  
prominent example is perhaps flat space where the metric certainly  
satisfies (\ref{confconds}) but there is no holographic argument in  
favor of identifying flat space with a confining gauge theory. We will  
nevertheless, devote some attention in section \ref{q} to the  
quantum corrections to  strings   
spinning in flat space  
for its historical and technical relevance to the topic of Regge  
trajectories.  
    
\section {Quantum corrected Regge trajectories\\ 
 for spinning strings   in flat space}   
\label{q}    
 
We begin with reviewing the general formalism of  semiclassical  
quantization  around  a given   
classical string configuration. The quantization procedure depends 
 on the string formulation used ( Polyakov  or  Nambu-Goto) and on the  
 gauge fixing.  Since we are particularly interested in spacetime 
 quantities like the energy and the angular momentum, it is important for 
 us to develop a general expression relating the spacetime quantities 
 to the worldsheet Hamiltonian (we follow the analysis of \cite{ft}).  
Let us consider the Polyakov formulation in the conformal gauge.   
In general,  for each of the string coordinates we turn on quantum fluctuations such that   
\be  
X^i(\s,\tau)=\bar  X^i(\s,\tau) + \delta X^i(\s,\tau),  
\ee   
where $\bar  X^i(\s,\tau)$ stands for the classical configuration. In  
particular, for the coordinates involved in defining energy and angular  
momentum  we have  
\be  
X^0 = e \tau + \delta X^0 \qquad \phi= e\o\tau + \delta \phi.  
\ee  
The Virasoro constraints:  
\be  
\eqlabel{viroconst}  
\begin{split}  
&g_{ij}\bigg[\pd_\tau X^i\, \pd_\tau X^j+\pd_\s X^i\, \pd_\s X^j  
\bigg]=0, \\  
&g_{ij}\pd_\tau X^i\, \pd_\s X^j=0,  
\end{split}  
\ee  
can be rewritten in the form of the requirement  
for the vanishing of the 2d Hamiltonian, namely  
\be  
{\cal H}(X^i) = {T_s\over 2}g_{ij}[   
\partial_\tau X^i  \partial_\tau X^j +  \partial_\s X^i  \partial_\s X^j] = 0.   
\ee 
 Now,  upon substituting $X^0= e\tau + \delta X^0$ we find that  
\be\label{gtt}   
T_s g_{00} \partial_\tau  X^0= {T_s \over 2 } e g_{00} + {1\over e}  
{\cal H}(\delta X^0 , X^i),  
\ee  
where we have replaced  $X^0$ by $\delta X^0 $ in the 2d Hamiltonian   
and $X^i$ now denotes the rest of the coordinates.  
We can now substitute (\ref{gtt}) into the expression for the space  
time energy. If we also   
explicitly introduce the classical configuration of $\phi$ and  
integrate we end up   
with the following expression for the space time energy  
\be  
E= \bar E + \o(J-\bar J) + {1\over e} \int {d\s}{\cal H}(\delta X^i)  
\label{vircobi}  
\ee  
 where $\bar E$ and $\bar J$ are the classical values of the energy and 
 angular momentum    
respectively and ${\cal H}(\delta X^i)$ is the 2d Hamiltonian where the  
original dependence on $X^i$  
is now replaced by the dependence on the fluctuations $\delta X^i$.  
What is left to be done is to  eliminate the dependence on  
$e$ and $w$ by using the   
expression for the classical angular momentum.      
The Virasoro constraint implies that   
\be  
H=0=J\dot \p+P_\r\dot \r- E\dot X^0 - L(P, q) +   
(d-3 \, {\rm massless}\,\,\, {\rm degrees} \,\,\, {\rm of}\,\,\,  
{\rm freedom}),   
\ee  
where   
$J=\int T_s\r^2\dot\p, \,\, P_\r=\int T_s\dot \r, \,\,E=\int T_s\dot  
X^0$. Explicitly, one can rearrange the Virasoro  
constraint as:  
\be  
\eqlabel{vird}  
\begin{split}  
eE -J&=\int d\sigma\bigg( H(d-3 \,  {\rm massless\,\,\, dof})  
+  
\frac {T_s}2(e^2-(\dot\d X^0)^2-(\d X^0{}')^2)   \\  
&+\frac {T_s}2(\bar\r+\delta \r)^2(-1+\dot{\d\p}^2+\d\phi'{}^2)  
+\frac {T_s}2 (\dot{\d\r}^2+(\bar\r+\d\r)'{}^2)\bigg)  \\  
&= e \bar E -  J +  
\int d\sigma  
(H(d-3 \,  {\rm massless\,\,\, dof})  \\  
&+ T_s \left(- \frac{\d \dot X^0{}^2}2-\frac{(\d X^0{}'){}^2}2 + \frac{\bar\r^2}2   
(\dot{\d\p}^2+\d\phi'{}^2)  
+\frac{\dot \d\r^2}2 +\frac{\dot \d\r'{}^2}2 - \frac12 \d\r^2 )+\dots  
\right) 
\end{split}  
\ee   
where the ellipsis  denotes  terms which are of higher order in fluctuations.  
On the right hand side of (\ref{vird}) we recognize the fluctuation Hamiltonian   
${\cal H}(\delta X^0,\delta\p, \d\r,\d\vec Z)$, where $\d\vec Z$ denotes   
collectively the $d-3$ massless degrees of freedom.    
The usefulness of this particular rewriting of  
the Virasoro constraints will 
become transparent in the covariant quantization of the Polyakov  
string. By taking the expectation value of (\ref{vird}) on the ground  
state where $J|\Psi\rangle = \bar J |\Psi\rangle$ one obtains  
\be  
\eqlabel{zero}  
e \Delta E = \int d\s <\Psi|{\cal H}|\Psi>.  
\ee  
The right hand side will compute the sum of the zero-point energies   
of all the degrees of freedom including the ghosts in covariant  
quantization. Thus, the right hand side of (\ref{zero}) is the sum of  
the zero-point energies of the   
physical degrees of freedom. This result formally obtained by assuming  
that the  
theory has been convariantly quantized turns out to be a   
physical one: it implies that in any quantization  
scheme the corrections to the spacetime energy will be given by the zero-point  
energies of the physical degrees of freedom on the worldsheet. In what follows in this  
section we will evaluate (\ref{zero}) by finding the spectrum of the  
physical degrees of freedom in a convenient gauge, instead  
of  using covariant quantization.     
\subsection{ Static open string: The L\"uscher term as a quantum correction}  

Prior to tackling the quadratic fluctuations of a spinning string in flat  
space let us recall the analysis for a static string in this background.  
The relevant classical solution is a straight line along, let say  
along $x_1$,  from $-L/2$ to $L/2$. In D flat dimensions the quantized action   
is simply that of $D-2$ massless fields associated with the $D-2$  
transverse directions. Assuming a time interval of $T$ and demanding that  
the eigenfunctions vanish on the boundary, the eigenvalues of the  
corresponding Laplacian are  
\be  
E_{n,m}= ({n\pi\over L})^2 + ({m\pi\over T})^2.  
\ee  
Therefore the free energy $F_B$ is given by (see appendix (\ref{fe}))  
\be  
-{2\over D-2}F_B = -{1\over 12} {T\over L}, 
\ee  
and we find the correction to the energy due to the   
bosonic fluctuations:  
\be  
F_b= (D-2){\pi\over 24} {T\over L},  \qquad \delta E= -{1\over T}F_B=  
-(D-2){\pi\over 24} {1\over L}.   
\ee  
This is an attractive L\"uscher term and we can see explicitly that  
its appearance is a one-loop effect from the string theory point of  
view.    Now, for a superstring one has to add the fermionic fluctuations.  
For a GS formulation in 10 flat space-time dimensions the fermionic action  
is  
\be  
S^{flat}_{fermion}= 2i T_s\int d\tau d\s \bar  
\theta\Gamma^\alpha\del_\alpha\theta  
\ee   
where $\theta$ are 10 -d Weyl Majorana spinor and $\Gamma^\alpha$ are the   
$SO(9,1)$ gamma matrices.  
Since the square of the fermionic operator  
$\Gamma^\alpha\del_\alpha=\nabla$  
and the eight transverse coordinate match the eight components of the Weyl  
Majorana spinor one finds that the L\"uscher term is canceled out since  
\be  
F= 8\times [-{1\over 2}\log \det \nabla+ \log \det(  
\Gamma^\alpha\del_\alpha)]=0.  
\ee  
Thus, in flat space the static string does not receive quantum  
corrections. A simple way to understand this situation is by verifying  
that a static open string in IIB is BPS. The analysis of supersymmetry is  
directly related to kappa-symmetry whereby we have  
\be  
\eqlabel{susy}  
X'^M\dot{X}^N\Gamma_{MN}\ep^*=\sqrt{-g}\ep.  
\ee  
For the static solution $X^0=\tau, X^1=\s=[-L/2,L/2]$. Therefore  
$\sqrt{-g}=1$ and \ref{susy} reduces to  
$\Gamma_{01}\ep^*=\ep$. Writing the complex spinor as $\ep=\ep_1+i\ep_2$  
and using the $\Gamma_{01}^2=1$ we see that we can arrange for  
$(1-\Gamma_{01})\ep_1=0$ which shows that the solution preserves half  
the supersymmetries.   
      
\subsection{Regge trajectories: The intercept as a quantum correction}  
 
We now proceed to study the quantum fluctuations up to quadratic order  
around the classical solution describing a spinning string.   
For pedagogical reasons we derive the spectrum of the quantum   
fluctuations by performing a semi-classical expansion around the  
spinning string solution using both the Nambu-Goto action and the   
Polyakov action. Our treatment allows to clearly visualize some of the  
shortcomings of these methods.      
Let us begin by considering the Nambu-Goto action of a bosonic string  
in a flat target-space background  
\bea  
S={T_s}\int&\bigg|&(-(\dot X^0)^2+ \dot \r^2 + \r^2\dot\phi^2+\dot{\vec 
    z}\dot{\vec z})(-(X^0{}')^2+\r'^2+\r^2\phi'\,{}^2 +  
\vec{z}\,{}' \vec{z}\,{}')\nonumber\\ 
&-&(-\dot X^0 X^0{}' + \dot \r \r' +  
\r^2 \dot \phi \phi'+  
\dot{\vec z}\vec{z}\,{}')^2\bigg|^{\frac 12}.  
\label{nambugoto}  
\eea  
Denoting by $\r, \phi$ the polar coordinates which parametrize the   
the plane where the string motion is confined $(ds^2 =-d(X^0){}^2 + d\r^2  
+\r^2\, d\phi^2+\ldots)$,  
we expand around the classical solution  
\be  
\eqlabel{polarsol}  
\bar X^0=e\tau~~~~;~~~~\bar \r=e \sin(\s)~~~~;~~~~ \bar\phi=e\omega\tau.  
\ee  
In what follows for simplicity we will  sometimes set $e\omega=1$.  
By  using diffeomorphism   
invariance to gauge-fix $X^0=\bar X^0, \r=\bar \r$, one obtains   
\be  
S_{NG}=T_s \left(\int \sqrt{e^4 \cos^4(\s)}+\frac{e^2\tan^2(\s)}2(-\d\dot\p^2   
+\d\p'\,{}^2) -\frac 12(\d\dot{\vec z}\d\dot{\vec z}-  
\d\vec{z}\,{}'\d\vec{z}\,{}')\right),  
\ee  
where $\delta\vec z(\s,t)$ denote the other $d-3$ fluctuations. To  
cast  the action for the fluctuation $\delta\phi$ in the form  
of a standard kinetic term plus a potential we make the field redefinition  
\be  
\eqlabel{redefbo}  
\d\tilde \p=\d\p\,\, \tan(\s),  
\ee  
and find  
\begin{equation}  
S_{NG}= T_s \left(\int e^2 \cos^2(\s)+{e^2\over 2}(-\d\dot{\tilde{\phi}}^2   
+\d\tilde{\phi}'\,{}^2) +{e^2\over 2} {2\over \cos(\s)^2}(\delta\tilde\phi)^2  
-\frac 12(\d\dot{\vec z}\d\dot{\vec z}-  
\d\vec{z}\,{}'\d\vec{z}\,{}')\right).  
\end{equation}  
To conclude, the Nambu-Goto action when expanded around the spinning string   
solution describes one massive fluctuation ($\d\tilde\p$)  
\be  
\eqlabel{massng}  
\left(\pd_\s^2 -\pd_\tau^2 -{2\over \cos^2(\s)}  
\right)\d\tilde \phi(\s)=0.  
\ee  
with a $\sigma$-dependent mass square $2/\cos^2(\sigma)$  
and $d-3$ massless fluctuations ($\d\vec z$).   One can show that the   
expansion of the $\delta\tilde \phi$ fluctuation in eigenmodes is given by  
\bea\label{eigenmode}  
\delta \tilde \phi (\sigma, \tau)& =&   
\tilde\phi_{nm}\chi_{nm}(\s,\tau)\nonumber\\  
&\equiv& \sum_{n}\frac{\exp(i n\tau)}{\sqrt{2\pi}}\bigg(  
\sum_{m=odd}\tilde\phi_{nm}   
\frac1{\sqrt{\pi(m^2-1)}}  
(m \sin(m\s) - \cos(m\s)\tan\s)\nonumber\\  
&+&\sum_{m=even} \tilde\phi_{nm}  
\frac1{\sqrt{\pi(m^2-1)}} (m\cos(m\s)+\sin(m\s)\tan\s)\bigg)\label{Fmd}  
\eea   
where $\chi_{nm}$ are normalized eigenfunctions of the differential operator  
\be  
\begin{split}  
&(\partial_\tau^2 -\partial_\s^2+\frac 2{\cos^2\s})\chi_{nm}(\s,\tau)=  
(m^2-n^2)\chi_{nm}(\s,\tau)\equiv \lambda_{nm}  
\chi_{nm}(\s,\tau)\label{chieqn},\\  
&\int_{-\pi}^\pi d\sigma~\chi_{nm}\chi_{n'm'}=e^{i \,(n+n')\tau}  
\delta_{m-m'}.  
\end{split}  
\ee  
The partition function of this massive mode will therefore be very  
similar to that of a massless fluctuation.   Certainly the eigenvalues  
as read off from (\ref{chieqn}) are the same as for a massless mode with  
standard kinetic term and standard Fourier mode decomposition. However,   
there is one difference between the spectrum of the $\d\tilde\phi$   
fluctuation and the spectrum of a massless fluctuation.   
Namely, the former has one less mode than the latter since,    
as can be seen from (\ref{eigenmode}),    the eigenmode with $m=1$ vanishes.
In such cases one looks for a special solution for this mode. Indeed, 
it is easy to  realize that the corresponding 
 equation $( -\partial_\s^2+\frac 2{\cos^2\s})\chi_{n \, m=1}(\s)=  
\chi_{n \,m=1}(\s)$ is solved by $\chi_{n \,m=1}= {1\over \cos \s}$.
However, this special eigenmode   
is not $\sigma$-normalizable and hence cannot be counted. 
 This peculiar situation leads to  a different answer for the energy.   
For instance, the zero mode energy calculation done at the beginning of this   
section becomes  
\be  
\sum_{\stackrel{n,m\in {\bf Z}}{m\neq 1}} \log E_{n,m}=  
\sum_{n,m\in{\bf Z}} \log E_{n,m}-\sum_{n\in{\bf Z}} \log E_{n,m=1}  
=\sum_n n -1=\zeta(-1)-1  
\ee  
In the above we have taken the simplified values   
$T=2\pi$,   
$L=2\pi$ since this is the periodicity that has been implicitly assumed  
in the Fourier mode decomposition (\ref{Fmd}).It is easy to restored  
the $T$ and $L$ dependence using the analysis of appendix (\ref{fe}).    
Let us see how the spacetime energy and angular momentum operators are expressed in terms  
of oscillators. The {\it target-space} energy is the canonical conjugate  
variable to $X^0$,   
while the angular momentum is the canonical conjugate variable to   
$\phi$.   Using the Nambu-Goto action (\ref{nambugoto})   
we derive that the energy and angular momentum, respectively, are   
\bea  
E&=& \int d\sigma \frac{\delta S}{\delta \dot X^0}\nonumber\\  
&=&T_s \int d\sigma \bigg[ {\dot X^0 (-X^0{}'^2+\r'^2+\r^2\phi'\,{}^2 +  
\vec z' \vec z')-X^0{}' (-(\dot X^0){}^2+ \dot \r^2 + \r^2\dot\phi^2 
+\dot{\vec 
    z}\dot{\vec z})} \bigg]
\nonumber\\ 
&\times &\bigg|(-(\dot X^0){}^2+ \dot \r^2 + \r^2\dot\phi^2 
+\dot{\vec 
    z}\dot{\vec z})(-(X^0){}'^2+\r'^2+\r^2\phi'\,{}^2 +  
\vec{z}\,{}' \vec z\,{}')\nonumber\\ 
&&-(-\dot X^0 X^0{}' + \dot \r \r' +  
\r^2 \dot \phi \phi'+  
\dot{\vec z}\vec z\,{}')^2\bigg|^{\frac 12},\nonumber\\  
\\  
J &=& \int d\sigma \frac{\delta S}{\delta \dot \p}  
\nonumber\\  
&=&T_s \int d\sigma \bigg[ {(-\r^2 \dot \p)   
(-(X^0){}'^2+\r'^2+\r^2\phi'\,{}^2 +  
\vec z\,{}' \vec z\,{}')+\r^2\p' (-(\dot X^0){}^2+ \dot \r^2 +  
\r^2\dot\phi^2+\dot{\vec 
    z}\dot{\vec z})}\bigg]\nonumber\\ 
&\times & \bigg|(-(\dot X^0){}^2+ \dot \r^2 + \r^2\dot\phi^2+\dot{\vec 
    z}\dot{\vec z})(-(X^0){}'^2+\r'^2+\r^2\phi'^2 +  
\vec z\,{}' \vec z\,{}')\nonumber\\ 
&&-(-\dot X^0 X^0{}' + \dot \r \r' + \r^2 \dot \phi \phi'+  
\dot{\vec z}\vec z\,{}')^2\bigg|^{\frac 12}. 
\eea  
Substituting the gauge choice $X^0=e\tau, \r=e\sin\s$ and expanding  
in fluctuations up to second order, one finds  
\bea  
E&=&T_s \int d\sigma  
\frac{1}{2e\cos^4\s}\bigg(\dot\p^2(3e^2-5e^2\cos^2\s+2e^2\cos^4\s)  
+\p'\,{}^2(-e^2+3 e^2\cos^2\s-2e^2\cos^4\s)\nonumber\\  
&+&\dot {\vec z}^2\cos^2\s+\vec z'\,{}^2 \cos^2\s\bigg),\\  
J&=&-T_s \int d\sigma\frac{\sin^2\s}{\cos^4\s}\bigg(-3e^2\sin^2\s\,\dot\p^2-e^2\sin^2\s\p'\,{}^2  
-\dot {\vec z}{}^2\cos^2\s-\vec z'\,{}^2 \cos^2\s\bigg).  
\eea  
As announced in the introduction to the section (\ref{q}),   
the difference $e E-J$ (which in the   
Nambu-Goto bosonic string turns out to be equal to the Hamiltonian of the   
physical degrees of freedom) when evaluated on the ground state   
yields  
\bea  
e(E-\bar E) &=& T_s\int d\s\langle\Psi|  
\frac 12(\dot{\vec z}^2+\vec z'{}^2 + e^2\tan^2\s(\dot\p^2+\p'{}^2)  
|\Psi\rangle   
\nonumber\\  
&=& \frac{\pi}2 [(D-3)\sum_{n>0} n + (\sum_{n>0} n -1)]\nonumber\\   
&=& \frac{\pi}2 \bigg(-\frac {D-2}{12} - 1 \bigg).  
\eea  
Note that the quantum correction  $E-\bar E$ can be interpreted as a Luscher term since it it proportional to ${1\over e}$ where $e$ is the length of the string.  
Using the classical values of $J$ and $E$ one obtains the quantum   
corrected Regge trajectory  
\be  
\eqlabel{regcor}  
J\approx {1\over 2}\a' \, E^2  + \frac \pi 2\left({D-2\over 12}+1\right)\equiv 
{1\over 2}\a' \, t + \frac{\pi}2 \left({D-2\over 12}+1\right) + O({1\over \a' t}).   
\ee  
Let us discuss the validity of the above result.   
The Nambu-Goto action is not free even in flat space since the   
expansion of the square root contains terms higher order in fluctuations. We  
have, naturally used a semiclassical approximation by truncating the  
action to terms quadratic in fluctuations. The approximation relies  
on the fact that higher order interaction terms are suppressed by  
powers of the energy of the state $1/E$. This is the reason why we  
neglect terms that are order $E^{-2}$ in our final expression for the  
quantum corrected Regge trajectories  (\ref{regcor}).  The net effect   
of the one loop quantum corrections on the Regge trajectory is to provide a   
{\it positive intercept}. A nonvanishing and positive intercept is a 
generic characteristic of Regge trajectories for mesons, but more 
importantly for the soft Pomeron \cite{pomeron}.  
\subsubsection{Fermions and the Regge intercept}  
The above discussion can be generalized to the case of 10-d  
Green-Schwarz superstring  
\bea  
S&=& \frac{T_S}2 \int d^2\sigma \sqrt{\gamma} \gamma^{\alpha\beta}  
\partial_\a X^\mu \partial_\b X_\mu   
\nonumber\\  
&+&  iT_s\int d^2\sigma (\sqrt{\gamma} \gamma^{\alpha\beta}\delta_{IJ}-  
\epsilon^{\alpha\beta} (\tau_3)_{IJ})   
\bar\theta^I\rho_\alpha\nabla_\beta\theta^J,  
\eea  
where $\alpha,\beta$ are worldsheet indices, $\gamma_{\alpha\beta}$  
is the worldsheet metric and $\gamma=\det (-\gamma_{\a\b})$.  
The matrices $\rho_\alpha$ are given by  
\be  
\rho_\a =e^m_\mu \partial_\a X^\mu \Gamma_m,  
\ee  
where $m$ is a flat target-space index and $\Gamma^m$ are the ten dimensional  
Dirac matrices.   
The derivatives $\nabla_\a$ are defined as  
\be  
\nabla_\a=\partial_\a +\frac 14 \Omega_\mu^{mn}\Gamma_{mn}\partial_\a X^\mu.  
\ee  
Finally, $\tau_3$ is one of the 2 dimensional Pauli matrices  
$\tau_3 = {\rm diag}(1, -1)$.Let us   
begin by gauge-fixing the kappa symmetry by choosing $\theta^1=\theta^2  
\equiv\theta$,   
such that from the beginning the only fermionic degrees of freedom   
are physical.   There are different ways to fix the conformal invariance but in  
accordance with our Nambu-Goto analysis we choose to integrate the  
world-sheet metric using its equation of motion.   
The effect of this algebraic operation is to remove the conformal  
invariance of the GS action, and the resulting action will be   
\be  
S=T_s \int d^2\sigma \sqrt{\det(\partial_\a X^\mu \nabla_\b X_\mu  
+\bar\theta \rho_\a \nabla_\b \theta}).  
\ee  
Next, by expanding to quadratic order in fluctuations around the classical  
configuration of a spinning string we find   
\be  
S= S_{NG}+ i\frac 12 T_s \int d^2\sigma   
\bar\theta (e\Gamma^0 - \bar\rho \Gamma^\phi )\dot\theta + \frac 12  
\bar\theta (e\Gamma^0 - \bar\rho \Gamma^\phi )  
\Gamma^{\phi\rho} \theta + \bar\theta   
(\bar\rho'\Gamma^\rho)\theta{}',   
\ee  
where $S_{NG}$ was given in (\ref{nambugoto}) and the Dirac matrices  
$\Gamma^0, \Gamma^\rho, \Gamma^\phi$ satisfy the usual Clifford algebra.  
The fermionic action can be rewritten with a standard kinetic term   
by making a unitary transformation similar to \cite{dgt}  
\be  
\eqlabel{rotation}  
\begin{split}  
\rho^0&=e\Gamma^0 - \bar\rho \Gamma^\phi\nonumber=  
e U \Gamma^0 \cos\sigma U^{-1},\\  
U&=\exp(-\frac 12 {\rm arccosh}(\frac{1}{\cos\sigma}) \Gamma^0\Gamma^\phi).  
\end{split}  
\ee  
Note that we can equally write  
\be  
\rho^1=e U \Gamma^\rho e\cos\sigma U^{-1}.  
\ee  
It is then natural to define new fermionic fluctuations  
\be  
\eqlabel{redef}  
\Psi(\s,\tau)  
=\sqrt{\cos\s} U^{-1}(\sigma)\theta(\s,\tau).   
\ee  
In terms of these new variables the fermionic action becomes   
\bea  
S_F&=&i \frac 12 e T_s   
\int d^2\sigma \left(\bar\Psi (\Gamma^0\partial_\tau+\Gamma^\rho  
\partial_\sigma) \Psi + \bar\Psi \Gamma^\rho U\sqrt{\cos\sigma}  
\partial_\sigma (U^{-1}\frac{1}{\sqrt{\cos\sigma}})\Psi\right.  
\nonumber\\  
&+&\left. \frac 12 \bar\Psi \Gamma^0\Gamma^{\p\rho}  
U^2 \Psi \right)\nonumber\\  
&=&i T_s \int d^2\sigma \left(\bar\Psi (\Gamma^0\partial_\tau+\Gamma^\rho  
\partial_\sigma) \Psi +\frac 1{\cos\sigma} \bar\Psi\Gamma^\rho\Gamma^0\Gamma^\p\Psi\right).  
\eea  
Choosing a specific realization of the Clifford algebra for the  
Dirac matrices that appear in the action  
$\Gamma^0=i\tau_2 \otimes \id_2\otimes \id_4,\,\,   
\Gamma^\rho=\tau_1\otimes \id_2\otimes\id_4, \,\,\Gamma^\phi=\tau_3\otimes\tau_1  
\otimes \id_4$,  
and decomposing the 16-component spinor $\Psi$ into four 4-component spinors  
$\psi^i=(\psi_1^i, \psi^i_2, \psi^i_3, \psi^i_4)$, $i=1,\dots 4$, 
 one observes that the Dirac   
equations obeyed by each of these four spinors split into two sets  
of coupled first order differential equations for the components  
$\psi^i_2, \psi^i_3$ and $\psi^i_1, \psi^i_4$ respectively. The  
equations satisfied by the pair $(\psi_2, \psi_3)$ are    
\bea  
&&(\partial_\tau+\partial_\sigma)\psi_2^i +\frac 1{\cos\sigma}  
(-\psi^i_3)=0,\\  
&&(-\partial_\tau+\partial_\sigma)\psi_3^i +\frac 1{\cos\sigma}  
(-\psi^i_2)=0.  
\eea  
Similar equations are satisfied by $(\psi^i_1, \psi^i_4)$.  
Thus, the mass of the eight fermionic fluctuations is $1/\cos^2\s$.
Note that this mass term is identical to the one found for the bosonic
fluctuation $\d\tilde \p$ (\ref{massng}). However, in the bosonic sector 
only one mode is massive and the rest seven physical modes are massless,
whereas there are eight massive fermionic modes. 
Altogether the contribution of the fermions to the intercept is $-{13\over 3}\pi$. Since the bosonic contribution is ${5\over 6}\pi$ the total intercept is
$-{7\over 2}\pi$. Notice that unlike for the bosonic string,
 the superstring  admits  a negative intercept.
One remark is in order. If from some reason, which at present we cannot find,
the $m=1$ mode discussed above can be rescued, then all the bosonic and 
all the fermionic modes are massless and there is an exact cancellation between them which yields a vanishing intercept. As we will show in the next subsection,
this is indeed the case in the covariant formulation.

\subsection{Covariant treatment of fluctuations: No corrections to   
Regge trajectories}  
\label{cov}    
 
The analysis in the previous section gives us a very direct way to  
built a solid intuition for what the quantum corrections should be  
like. There is,  
however, a very subtle point that only the covariant approach can  
clarify. Namely the implications of the field redefinitions used in  
both approaches on the quantum measure. Note that we have redefined  
both the bosonic field (\ref{redefbo}) and the fermionic fields  
(\ref{redef}). Normally, as has been the case for some static  
configurations in $AdS_5\times S^5$ \cite{difference}, the field  
redefinition   
is harmless as far as a quantum  
Jacobian for the measure is concerned. In particular, as explained in  
\cite{dgt}, some of the divergences are subleading and can be ignored.    
In our case, however, we will  
see that the situation is different. A careful analysis of the  
fluctuations that takes into consideration issues of the measure  
yields a slightly different result from the naive Nambu-Goto  
approach.   
 
 
We will analyze the fluctuations in the context of the  
Polyakov action in both  a path integral approach 
 as well as  a canonical quantization procedure. In both cases we use Cartesian coordinates.

\subsubsection { Path integral quantization} 
  The most transparent parametrization of the bosonic  
fluctuations uses   
the symmetries of the problem. Henceforth,   
we make the standard decomposition into background fields and quantum   
fluctuations:  
\be  
X^m = \bar X^m + \delta X^m.     
\ee  
Next we choose the conformal gauge  
\be  
\eqlabel{confg}  
\g_{\a \b}=\partial_\a \bar X^m \partial_\b \bar   
X^n \eta_{mn} = e^2 \cos^2\s \;\eta_{\a\b}=\sqrt{\g}\eta_{\a\b},  
\ee  
by identifying the world-sheet metric with the induced metric by the   
target-space background configuration.   
The bosonic fluctuations are world-sheet scalars,  
and their measure   
is given by  
\be  
||\delta X^m||^2 = T_s\int d\s d\tau \sqrt{\g} \delta X^m \delta X^m.  
\ee  
In the Polyakov string formulation, the bosonic fluctuations 
when expressed in Cartesian coordinates are decoupled  
and their action is simply 
\be 
S_B=\frac {T_s}2 \int d\tau d\sigma\; \sqrt{\gamma}\gamma^{\alpha\beta} 
\partial_\alpha \delta X^m \partial_\beta \delta X_m 
=\frac {T_s}2 \int d\tau d\sigma\; 
\eta^{\alpha\beta}\partial_\alpha \delta X^m \partial_\beta \delta X_m. 
\ee 
Therefore, by integrating them out in path integral one obtains  
\be   
\int DX \;e^{S_{cls}} \exp(T_s\int d\s d\tau \sqrt{\g} \g^{\a\b}  
\partial_\a \delta X^m  
\partial_\b \delta X_m) =e^{S_{cls}}\det(\Delta_{\g})^{-d/2},  
\ee  
where $d$ is the number of target-space dimensions, and   
\be  
\Delta_\g= {1\over \sqrt{\g}}  \partial_\a \g^{\a\b} \sqrt{\g}\partial_\b,  
\ee  
is the two-dimensional world-sheet Laplace operator.    
 
The contribution of the reparametrization ghosts to the  
partition function   
(omitting the zero-mode contributions) is  
\be  
\int Db Dc \; \exp({T_s\int d\sigma d\tau \sqrt{\g} \; b_{\a\b}\g^{\a\g}  
\g^{\b\d}   
\nabla_\g c_\d})=  
\det(P_1 P_1^\dagger)^{1/2}  
\ee  
where $P_1$ is the differential operator that maps vectors into symmetric  
traceless two-tensors. On a genus one world-sheet one has  
\be  
\det(P_1 P_1^\dagger)^{1/2}=\det(\Delta_{\g})  
\ee  
and thus the contribution of the two longitudinal bosonic degrees of freedom   
is canceled by ghosts.    
 
Finally we are left with the fermionic degrees of freedom.  
The starting point is the Green-Schwarz superstring, with the fermionic   
degrees of freedom reduced to the physical ones by choice of kappa-gauge.  
As before, we choose to identify $\theta^1=\theta^2\equiv \theta$.  
Expanding 
the Polyakov action to second order in the fermionic fluctuations $\theta$ 
yields the following action: 
\bea 
S_F&=&\frac i2  
T_s \int d\tau d\sigma\;\sqrt{\gamma} 
\gamma^{\a\b}\partial_\a \bar X^\mu \bar\theta 
\Gamma_\mu \partial_\beta \theta 
\nonumber\\ 
&=&\frac i2 T_s\int d\tau d\sigma\;\bigg[e\bigg(\bar\theta 
(\Gamma^0 + (\Gamma^1\cos\tau-\Gamma^2\sin\tau)\sin\s)\dot\theta 
\nonumber\\ 
&+&\bar\theta( \Gamma^1\sin\tau+\Gamma^2\cos\tau)\cos\s \theta' \bigg) 
\bigg]. 
\eea  
Using that through unitary transformations 
\bea 
U_{01}&=&\exp(-\frac 12{\rm arccosh}(\frac 1{\cos\sigma})\Gamma^0\Gamma^1)\\ 
U_{12}&=&\exp(\frac 12\tau\Gamma^1\Gamma^2) 
\eea 
we can rewrite  
\bea 
\Gamma^0 + (\Gamma^1\cos\tau-\Gamma^2\sin\tau)\sin\s&=&  
\cos\sigma U_{12} U_{01}\Gamma^0 
U_{01}^{-1} U_{12}^{-1}\\ 
(\Gamma^1\sin\tau+\Gamma^2\cos\tau)\cos\s&=&\cos\sigma U_{12} U_{01}\Gamma^2 
U_{01}^{-1} U_{12}^{-1}, 
\eea 
the action simplifies, provided that we make the field redefinition 
$\Psi=U_{01}^{-1} U_{12}^{-1}\theta$, to 
\be\label{GSCar} 
S_F=\frac i2  
T_s \int d\tau d\sigma\; \cos\sigma \bar\Psi(\Gamma^0\partial_\tau+\Gamma^2\partial_\sigma)\Psi. 
\ee 
Integrating out the fermionic fluctuations, which are world-sheet scalars,  
with the path integral measure  
\be  
|| \bar \theta \theta || =T_s \int d\sigma d\tau   
\sqrt{\g}  \bar \theta \theta,  
\ee  
one finds their contribution to the partition function  
\bea  
\label{ferdet}  
&&\int D\theta \;\exp({\int d\sigma d\tau \sqrt{\g}   
\bar\theta \g^{\a\b}\partial_\a \bar X^m \Gamma_m \partial_\b \theta})  
= \det(\g^{\a\b} \partial_\a \bar X^m \Gamma_m \partial_\b)^8 \nonumber\\  
&=& \det(\g^{\a\b} \partial_\a \bar X^m \Gamma_m \partial_\b  
(\g^{\g\d} \partial_\g \bar X^n \Gamma_n \partial_\d))^4  
=\det(\sqrt{\g}^{-1} \eta^{\a\b} \eta^{\g\d}  
\partial_\a \bar X^m \partial_\b (\sqrt{\g}^{-1}\partial_\g \bar X_m   
\partial_\d))^4\nonumber\\  
&=& \det(\g^{\a\b} \partial_a\partial_b)^4.  
\eea  
In the last step we used repeatedly equation (\ref{confg}) and the  
fact that partial derivatives commute.  To explicitly evaluate  
(\ref{ferdet}) we use the fact that the determinants of two conformally  
related operators are also related by a simple relation \cite{dgt} (see  
also \cite{conformaldet}). In particular the fermionic determinant 
 (\ref{ferdet}) can be reexpressed as 
\be  
\eqlabel{fermion}  
\begin{split}  
&\ln(\det(\g^{-1/2} \delta^{\a\b}  
\partial_\a \partial_\b))-\ln(\det(\d^{\a\b}\partial_\a \partial_\b))\\  
&= -\frac 1{4\pi}\int_0^{2\pi}  
d \sigma \int_0^T d\tau \frac 1{12}\d^{\a\b}\partial_\a\ln{\sqrt{\g}}  
\partial_\b \ln{\sqrt \g}\\  
&= -\frac T6  
\end{split}  
\ee  
The evaluation of the similar determinant for the bosons follows the  
same pattern:   
\be  
\eqlabel{boson}  
\begin{split}  
&\ln(\det(\Delta_\g)-\ln(\det(\delta^{\a\b}\partial_\a\partial_\b)\\  
&= -\frac 1{12\pi}\int_0^{2\pi}  
d \sigma \int_0^T d\tau \d^{\a\b}\partial_\a\ln \g^{1/4}  
\,\,\,\partial_\b \ln \g^{1/4}\\  
&=-\frac T6  
\end{split}  
\ee  
One crucial element in this analysis that differs  
from the naive Nambu-Goto approach is that the path integral approach  
captures some divergences embodied in the $-T/6$ terms. If desired, we  
can explicitly evaluate all the determinants presented above  
(\ref{fermion}) and (\ref{boson}) since the right hand side is  
completely explicit and the second term in the left hand sides has  
been explicitly evaluated in appendix \ref{fe}.  
 
 
\subsubsection{ Canonical quantization} 
It is obvious that the $D-3$ ( seven  in our case) $\vec z$ coordinates 
are decoupled from the classical configuration and hence in flat space time 
the fluctuations in these coordinates are free massless modes. So from here on  we 
discuss only the rest of the coordinates.  
In the Cartesian coordinates basis  the decomposition of the fields to the classical configurations and the quantum fluctuations take the form 
\bea 
X^0 &=&  e\tau + \delta X^0 \nonumber\\ 
X^1 &=& e \cos\tau \sin\s  + \delta X^1 \nonumber\\ 
X^2 &=& e \sin\tau \sin\s  + \delta X^2 \nonumber\\ 
\eea 
Upon  inserting  these configuration into the Polyakov action 
in the covariant gauge, we easily find from  
the equations of motions  that the fluctuations are all massless, namely, obey 
\be 
\left(\pd_\s^2 -\pd_\tau^2   
\right)\delta X^i =0  
\ee 
in particular for $i= 0,1,2$. 
The Virasoro constraint takes the form of (\ref{vircobi}) where  
${\cal H}(\delta X^i)$ is the 2d Hamiltonian density  of the 10 free massless bosonic modes. 
In the covariant gauge, however, as was discussed above,  
one has to take into account the contribution 
of the reparametrization ghosts. It is well know that the latter eliminate the  
longitudinal modes and hence the final form of the bosonic part of the Virasoro constraint is  
\be 
e(E-\bar E)= -\pi{D-2\over 24}= -{\pi\over 3} 
\ee 
Note that in deriving this result in the Polyakov formulation we have not taken  
a quadratic approximation, but rather the full action. The only approximation 
made here is setting in (\ref{vircobi}) the value of $J$ to be equal to $\bar J$. 
 
Next we discuss the fermionic contribution to the energy $E$.  
It was shown above in (\ref{GSCar}) that the relevant operator for the  
fermionic determinant is $(\Gamma^0\partial_\tau+\Gamma^2\partial_\sigma)$. 
This means that unlike in the polar coordinates formulation, here there is  
no mass term and  
the fermionic modes, just as the bosonic ones,  
are  free and massless. 
This of course implies  
 that  they  contribute to the Virasoro constraint the same contribution as the  
bosonic modes but with an opposite sign namely   ${\pi \over 3}$. 
Hence, as for the path integral  approach,  also
in  the canonical quantization  the bosonic and  
fermionic contributions to the intercept cancel each other.  
 

The main conclusion of this section is upon using the Polyakov formulation
in the Cartesian coordinates we observe 
 that for the superstring  
spinning in flat space there are no quantum corrections to the  
intercept which  therefore remains zero.    
Hence it is  
clear  from both the path integral and the canonical quantization  
that the semiclassical partition   
function of the GS superstring whose classical configuration describes a   
string which spins in flat space is trivial, being identical with 1.  

To summarize, we have presented two types of calculations of 
the quantum fluctuations around the classical spinning string in flat space-time. In the first
we used 
  the  canonical quantization of the NG formulation in polar coordinates and in the second 
the Polyakov formulation in Cartesian coordinates both in a path-integral as well as a canonical quantizations.
Recall  that 
 in the GS framework the NG and Polyakov  fermionic actions are the same.
 In the NG formulation of the bosonic modes we have truncated the action to 
include only quadratic terms whereas in Polyakov's formulation we used the 
exact expression with no truncation since it is quadratic. 
The outcome of the two types of evaluations is different. In the first we find
a non-trivial intercept whereas in the second the intercept vanishes. 
Now since we do not see any loopholes in the second approach we believe that the correct result is that there is no intercept. As it stands the results that 
follow from the NG approach are in contradiction with this conclusion. 
At this point there are two options: (i) that indeed the NG approach leads to 
a different result (ii) that form a reason that we do not understand at present
the $m=1$ modes discussed above are not missing from the spectrum of eigenmodes
and hence both the bosonic and fermionic modes admit a massless spectrum and 
the intercept vanishes. We leave the investigation of this open question for 
a future investigation.

\subsection {Supersymmetry}   
 
In this subsection we show that, despite the one-loop cancellation just 
presented, the string configuration we are  
considering is not supersymmetric.  
The supersymmetry condition follows directly from the  
kappa-symmetry variation of the action (\ref{susy}). In the particular  
background we are considering one has   
\be  
\begin{split}  
X'^M\dot{X}^N\Gamma_{MN}\ep^*&=\sqrt{-\g}\ep \\  
\rho'\dot X^0\Gamma_{\r 0}+\rho' \dot\phi \Gamma_{\rho \phi} \ep^*&= 
e^2\cos^2  
\s \ep \\  
\cos\s \Gamma_{\r}\left(\Gamma_{0}+\Gamma_{\r}\right)\ep^*&=\cos^2\s \ep\\  
\G^{\hat{\rho}}\left(\G^{\hat{0}}-\sin\s 
  \G^{\hat{\phi}}\right)\ep^*&=-\cos\s \ep.  
\end{split}  
\ee  
Since the Killing spinor of spacetime is constant (flat space), the above expression 
implies that no supersymmetry is preserved by this solution. The result is also 
intuitively clear from the string theory point of view. Since the 
configuration has finite extension it does not belong in the 
supergravity part of the spectrum which is the sector where 
supersymmetric configurations are most likely to be found.  
Moreover, we know that this classical configuration describes massive 
string states.  Of course, 
some extended configurations are supersymmetric, like the straight 
string mentioned previously.  However, spinning strings  
typically break all supersymmetries due to the fact that the spin does 
not enter as a central charge in the supersymmetry algebra  and 
therefore can not support a BPS-like condition (see 
\cite{david}  for some recent considerations)\footnote{We thank 
D. Mateos for extensive discussions on issues of supersymmetry of 
spinning classical string configurations.}. The breaking of the 
supersymmetry makes the  origin of the one-loop  
cancellation obtained in the previous section unclear to us. We will 
simply point out that other nonsupersymmetric configurations, like the 
circular Wilson loop\footnote{See \cite{green} for an extensive discussion of 
the supersymmetry structure of the circular Wilson loop.}, enjoy very 
similar one-loop cancellations \cite{dgt}.

\section {Quadratic fluctuations in confining backgrounds}   
\label{quadra}  
 
Quadratic fluctuations of classical configurations in confining string  
backgrounds were analyzed for the static Wilson line configurations\cite{Kinar:1999xu}.  
The quantum fluctuations result in a typical L\"uscher term     
which introduces ${1\over L}$ corrections to the linear potential. These  
corrections carry information about the spectrum of the theory and can,  
in principle, be measured on the lattice. In this paper we compute  
quantum corrections which are definitely measurable as the intercept of  
Regge trajectories. Our aim is to determine the universal features of  
such corrections. To clarify the universality we conduct our analysis of  
the quadratic fluctuations for  the strings spinning in Sugra  
backgrounds dual to confining gauge theories. We consider the KS and MN  
backgrounds explicitly.     
\subsection{The Klebanov-Strassler background}  
\label{kssection}    
 
We begin by reviewing the KS background, which  
is obtained by considering a collection of  $N$  regular  
and $M$ fractional D3-branes in the geometry  of the deformed conifold  
\cite{ks}.  The 10-d metric is of the form:   \be  ds^2_{10} =   h^{-1/2}(\tau)    
dX_\mu dX^\mu  +  h^{1/2}(\tau) ds_6^2  
\ ,  \ee  where $ds_6^2$ is the metric of the deformed conifold  
\cite{candelas,mt}:   \be  
\label{mtmetric}  
ds_6^2 = {1\over 2}\varepsilon^{4/3} K(\tau)  \Bigg[ {1\over 3  
K^3(\tau)} (d\tau^2 + (g^5)^2)  +  \cosh^2 \left({\tau\over  
2}\right) [(g^3)^2 + (g^4)^2]  + \sinh^2 \left({\tau\over  
2}\right)  [(g^1)^2 + (g^2)^2] \Bigg].  
\ee  
where  
\be  
K(\tau)= { (\sinh (2\tau) -  
2\tau)^{1/3}\over 2^{1/3} \sinh \tau},  
\ee  
and  
\bea  
\label{forms}  
g^1 &=&  
{1\over \sqrt{2}}\big[- \sin\theta_1 d\phi_1  -\cos\psi\sin\theta_2  
d\phi_2 + \sin\psi d\theta_2\big] ,\nonumber \\  g^2 &=& {1\over  
\sqrt{2}}\big[ d\theta_1-  \sin\psi\sin\theta_2 d\phi_2-\cos\psi  
d\theta_2\big] , \nonumber \\  g^3 &=& {1\over \sqrt{2}} \big[-  
\sin\theta_1 d\phi_1+  \cos\psi\sin\theta_2 d\phi_2-\sin\psi d\theta_2  
\big],\nonumber \\  g^4 &=& {1\over \sqrt{2}} \big[ d\theta_1\ +  
\sin\psi\sin\theta_2 d\phi_2+\cos\psi d\theta_2\ \big],   \nonumber  
\\  g^5 &=& d\psi + \cos\theta_1 d\phi_1+ \cos\theta_2 d\phi_2.  
\eea  
The 3-form fields are:  
\begin{eqnarray}  
F_3 &=& {M\alpha'\over 2} \left \{g^5\wedge g^3\wedge g^4 + d [  
F(\tau)  (g^1\wedge g^3 + g^2\wedge g^4)]\right \} \nonumber \\  &=&  
{M\alpha'\over 2} \left \{g^5\wedge g^3\wedge g^4 (1- F)  + g^5\wedge  
g^1\wedge g^2 F \right. \nonumber \\  && \qquad \qquad \left. + F'  
d\tau\wedge  (g^1\wedge g^3 + g^2\wedge g^4) \right \}\ ,  
\end{eqnarray}  
and    
\begin{equation}  
\eqlabel{bfield}    
B_2 = {g_s M \alpha'\over 2} [f(\tau) g^1\wedge g^2  +  
k(\tau) g^3\wedge g^4 ]\ ,    
\ee  
\begin{equation}  
\eqlabel{h3}  
\begin{split}  
H_3 = dB_2 &= {g_s M \alpha'\over 2} \bigg[  d\tau\wedge (f'  
g^1\wedge g^2  +  k' g^3\wedge g^4)  \nonumber \\    
& \left. + {1\over  
2} (k-f)  g^5\wedge (g^1\wedge g^3 + g^2\wedge g^4) \right]\ .  
\end{split}  
\end{equation}  
The self-dual 5-form field strength is  decomposed as $\tilde F_5 =  
{\cal F}_5 + \star {\cal F}_5$, with    
\be    
{\cal F}_5 = B_2\wedge F_3  
= {g_s M^2 (\alpha')^2\over 4} \ell(\tau)  g^1\wedge g^2\wedge  
g^3\wedge g^4\wedge g^5\ ,    
\ee    
where    
\be    
\ell = f(1-F) + k F\ ,  
\ee    
and    
\be    
\star {\cal F}_5 = 4 g_s M^2 (\alpha')^2  
\varepsilon^{-8/3}  dx^0\wedge dx^1\wedge dx^2\wedge dx^3  \wedge  
d\tau {\ell(\tau)\over K^2 h^2 \sinh^2 (\tau)}\ .    
\ee    
The functions  
introduced in defining the form fields are:    
\bea  F(\tau) &=& {\sinh \tau -\tau\over 2\sinh\tau}\ ,  \nonumber \\    
f(\tau) &=&{\tau\coth\tau - 1\over 2\sinh\tau}(\cosh\tau-1) \ ,  \nonumber \\  
k(\tau) &=& {\tau\coth\tau - 1\over 2\sinh\tau}(\cosh\tau+1)  \ .  
\eea    
The equation for the warp factor is    
\be   
\label{firstgrav}  h' =  
- \alpha {f(1-F) + kF\over K^2 (\tau) \sinh^2 \tau}  \ ,    
\ee    
where  
\be    
\alpha =4 (g_s M \alpha')^2  \varepsilon^{-8/3}\ .    
\ee  
For large $\tau$ we impose the boundary condition that  $h$  
vanishes. The resulting integral expression for $h$ is    
\be  
\label{intsol}    
h(\tau) = \alpha { 2^{2/3}\over 4} I(\tau) =  (g_s  
M\alpha')^2 2^{2/3} \varepsilon^{-8/3} I(\tau)\ ,    
\ee    
where    
\be  
I(\tau) \equiv  \int_\tau^\infty d x {x\coth x-1\over \sinh^2 x}  
(\sinh (2x) - 2x)^{1/3}  \ .    
\ee    
The above integral has the  
following expansion in the IR:    
\be  I(\tau\to 0) \to a_0 - a_1  
\tau^2 + {\cal O}(\tau^4) \ ,    
\ee    
where $a_0\approx 0.71805$ and  
$a_1=2^{2/3}\, 3^{2/3}/18$. The absence  of a linear term in $\tau$  
reassures us that we are really expanding  around the end of space,  
where the Wilson loop will find it more favorable to arrange itself.       
\subsection{Quadratic fluctuations of the KS model}  
 
We consider the quadratic fluctuations and their influence on the  
Regge trajectories (\ref{Regge}).   
The full string theory in such  
backgrounds is not known. However, for a semiclassical treatment the  
sigma model action is needed only up to quadratic   terms and it is  
given by (we follow \cite{russo}):  
\be    
\eqlabel{action}  
\begin{split}  
S&={1\over 4\pi \a'}\int d\s_1 d\s_2 \sqrt{\g}\,  
\bigg[(g_{\m\n}\g^{\a\b}+b_{\m\n}\epsilon^{\a\b})    
\partial_\a X^\m \partial_\b X^\n  \\    
&+i (\g^{\a\b} \delta_{IJ} -   \epsilon^{\a\b}    
(\rho_{3})_{IJ}) \partial_\a    X^{m } \bar \theta^I \Gamma_{m} D_\b    
\theta^J  \bigg],  
\end{split}      
\ee    
where  $\theta^I$ ($I$=1,2)  are the  two    
real positive chirality  10-d  MW spinors  and    
$D_b$ is the pullback to the world-sheet of the supergravity    
covariant derivative in the variation of    
the gravitino :    
\begin{equation}       
D_\a =  \partial_\a   + { 1 \over 4}    
\partial_\a X^\m \big[\     (\omega_{ m n \m}   - {1\over 2}  H_{ mn    
\mu} \rho_3) \Gamma^{ m  n}  +  ( { 1 \over  3!}    
F_{mnp} \Gamma^{mnp} \kappa_1  +   { 1 \over 2\cdot    
5!} F_{mnpqr}   \Gamma^{    
mnpqr}\rho_0 ) \Gamma_\m  \  \big]    \label{covderiv}  
\end{equation}    
where  the $\r_s$-matrices in the $I,J$ space  are the Pauli matrices    
$ \kappa_1 = \sigma_1$, $\rho_0 = i \sigma_2$, $\rho_3 =\sigma_3$.              
Let us first consider the  
metric. The part of the metric perpendicular to the world volume, which is   
the deformed conifold metric, does not enter in the classical  
solution which involves only world volume fields. Noting that the value  
$r_0$ of section \ref{stringconfine} is $\tau=0$, we expand the  
deformed conifold up to quadratic terms in the coordinates:   \begin{equation}  
\eqlabel{defcon}  
ds_6^2={\varepsilon^{4/3}\over 2^{2/3} 3^{1/3}}\bigg[{1\over 2} g_5^2  
+g_3^2+g_4^2+ {1\over 2} d\tau^2 +{\tau^2\over 4}(g_1^2+g_2^2)\bigg].  
\end{equation}  
Let us  further discuss the structure of this metric. It is known on  
very general grounds that the deformed conifold is a cone over a space  
that is topologically $S^3\times S^2$ \cite{candelas}. We can see  
that the $S^3$ roughly spanned by $(g_3,g_4,g_5)$ has finite size,  
while the $S^2$ spanned by $(g_1,g_2)$ shrinks to zero size at the  
apex of the deformed conifold. Note that the radius of the $S^3$ is  
given by the deformation parameter $\varepsilon$. What is less known  
is that up to the level of accuracy of (\ref{defcon}),  
$(g_3,g_4,g_5)$ spanned a space t  
that is not only topologically  but also geometrically an $S^3$ of  
radius $\sqrt{2}$ \cite{mt}. This fact  follows from the explicit  
construction of a matrix $T\in SU(2)$ such that   
\begin{equation}  
{1\over 2}{\rm Tr}\, dT^{\dagger}dT={1\over 2} g_5^2+g_3^2 +g_4^2.  
\end{equation}  
Because $T\in SU(2)$ the resulting metric is a round $S^3$ with radius  
$\sqrt{2}$. Thus, we introduce a standard parametrization of $S^3$ in terms of  
Euler angles $(\theta,\phi,\psi)$ and identify  
\begin{equation}  
{1\over 2} g_5^2+g_3^2 +g_4^2={1\over  
2}(d\theta^2+d\phi^2+d\psi^2+2\cos\theta \,\,d\phi \,d\psi).  
\end{equation}  
Expanding around the classical value of these coordinates  
$\phi=\psi=0, \,\,\theta =\pi/2$ we obtain essentially an $\mathbb{R}^3$  
which we parametrize by $(y_1,y_2,y_3)$. We end up with  
\begin{equation}  
{1\over 2} g_5^2+g_3^2 +g_4^2 \longrightarrow {1\over  
2}\big[dy_1^2+dy_2^2+dy_3^2\big].  
\end{equation}  
It must be the case then that $(g_1,g_2)$ spanned a space that is  
topologically an $S^2$. Because this $S^2$ space is fibered over $S^3$  
the combination $g_1^2+g_2^2$ will necessarily contain some of the  
coordinates $(\theta,\phi,\psi)$. However, up to quadratic terms we  
can neglect them\footnote{This situation is completely different from  
the discussion of \cite{gpss} where the precise structure of the  
fibration is crucial.}. Thus, to this level of accuracy, we can   
confidently identify $g_1^2+g_2^2$ with a round $S^2$ with a standard  
parametrization $(\bar\theta,\bar\phi)$. This round $S^2$ has  
coefficient  $\tau^2$ and   
combines with the $\tau$ direction  to give an $\mathbb{R}^3 $, which  
we choose to parametrize by $(\tau_1,\tau_2,\tau_3)$:  
\begin{equation}  
{1\over 2}\big[d\tau^2+{\tau^2\over  
2}(g_1^2+g_2^2)\big]\longrightarrow {1\over  
2}\big[d\tau_1^2+d\tau_2^2+d\tau_3^2 \big].  
\end{equation}  
To complete the bosonic part of the string action we need to consider the B-field (\ref{bfield})  
in this limit, that is, up to quadratic terms in the action  
(\ref{action}). The B-field contribution to the action at this order  
vanishes. A simple way to confirm this is by analyzing the structure  
of the NSNS  3-from 
 field strength (\ref{h3}). Let us consider the first term in 
 (\ref{h3}). Since near $\tau=0$  we have the following expansion 
 $f'\approx \tau^2/4 +{\cal O}(\tau^4)$ we see that this term is 
 proportional to the volume element of $\mathbb{R}^3$ defined by 
 $(\tau, g_1,g_2)$, thus in the new coordinates this term of the NSNS 
 3-form field strength is proportional to $d\tau_1\wedge d\tau_2\wedge 
 \tau_3$. In the action this term contributes in the form  
\be  
{1\over 2}{g_sM\a'\over 2}\,\,\tau_1\, \del_\a \tau_2 \del_\b \tau_3  
\,\,\epsilon^{\a\b},  
\ee  
which is third order in the action. Similarly, one can check that all  
other terms in the B-field contribute terms that are third and higher  
order and can, therefore, be neglected in the  approximation we are  
working. Note, that the situation is, again, different from  
\cite{gpss} where the B-field contributed to the quadratic string  
action. The reason being that if we take the light-cone gauge  
involving any of the coordinates $y_i$, then the above term becomes  
quadratic. Here, we do not anticipate taking a light-cone gauge that  
involves any of the six directions of the deformed conifold.      
Interestingly,  
there is an effective $\s$-dependent mass term for the  
$\tau$ directions. It arises as we expand in the GS action the  
warp factor up to quadratic order in $\tau$  
\be 
\frac 12 \int \sqrt\gamma \gamma^{\alpha\beta} ( 
g_{\tau\tau}(\bar X)\bigg|_{\tau=0} 
\partial_\alpha\tau^i \partial_\beta \tau_i 
+\frac 12  
\frac{\partial^2 g_{\mu\nu}(\bar X)}{\partial \tau^2}\bigg|_{\tau=0}  
\partial_\alpha \bar X^\mu \partial_\beta \bar X^\nu\; \tau^i\tau_i) 
\ee 
Using the conformal gauge 
$\gamma_{\alpha\beta}=\eta_{\alpha\beta} \sqrt{\g} $ 
with the choice of the Weyl factor such that the worldsheet metric is 
equal to the target-space induced metric 
\be 
\gamma_{\alpha\beta} 
= \partial_\alpha \bar X^\mu \partial_\beta \bar X^\nu g_{\mu\nu}(\bar X) 
\bigg|_{\tau=0} 
={1\over 2^{1/3}a_0^{1/2}}\,\, \frac{\varepsilon^{\frac 43}}{g_s M\alpha'} \,\, e^2\cos^2\sigma  \,\,\, 
\ee 
we find the action for the three massive $\tau$ fluctuations 
\bea 
\frac 12 \int \sqrt{\g} \bigg(\sqrt{\g^{-1}} 
g_s M\alpha' {a_0^{1/2}\over 2^{4/3}3^{1/3}}  \;\eta^{\alpha\beta}\; 
\partial_\alpha \tau^i  
\partial_\beta \tau_i 
+\frac {a_1}{2 a_0} \;\tau^i \tau_i \bigg) 
\eea 
If one is willing to forget about the subtle issues of the path integral 
measure, then one would conclude that the $\tau^i$ fluctuations 
have a $\sigma$-dependent ``mass term''  
\be 
m_B^2= \frac{2\cdot 3^{1/3}\, 
  a_1}{a_0^{2}}\,\,\,\frac{\varepsilon^{4/3}}{g_s M \alpha'} \,\,\, 
\frac{e^2}{g_s M \alpha'}\,\,\,\cos^2 \s \equiv 2 m_0^2 e^2\cos^2\s. 
\ee 
Note that  $m_0$ is  proportional to the glueball mass  deduced from the 
dilaton spectrum in the KS background. Recall that the latter corresponds to  
a glueball of zero angular momentum.

\subsubsection*{Fermionic sector}   
 
Since the fermions are already quadratic in fluctuations given that  
their classical value is zero we need to worry only about the value  
of the field strength  at the point $\tau=0$. This means that   
effectively  $H_{NS}$ vanishes and the RR 3-form $F_3$ is completely  
directed  along the $S^3$ directions:  
\be  
F_3= \frac{M\alpha'}{2}\,\, g_5\wedge g_3\wedge g_4  
\ee

\subsection {The Maldacena-N\`u\~nez background}   
\label{mnsection}    
 
The MN background whose IR regime is associated with ${\cal N}=1$  
SYM theory is that of a large number of D5 branes wrapping an  
$S^2$. To be more precise: (i) the dual field theory to this  
SUGRA background is the ${\cal N}=1$ SYM contaminated with KK  
modes which cannot be de--coupled from the IR dynamics, (ii) the  
IR regime is described by the SUGRA in the vicinity of the origin  
where the $S^2$ shrinks to zero size.   The full MN SUGRA background  
includes the metric, the  dilaton and the  
RR three-form.   It can also be interpreted as an uplifting to ten  
dimensions a solution of seven dimensional gauged supergravity  
\cite{volkov}.  The metric and dilaton of the background are   
\bea    
ds^2&=&e^\p\bigg[dX^adX_a+\a'g_s    
N(d\tau^2+e^{2g(\tau)}(e_1^2+e_2^2)+{1\over 4} (e_3^2+e_4^2+e_5^2))\bigg],    
\nonumber \\    
e^{2\p}&=&e^{-2\p_0}{\sinh 2\tau\over 2e^{g(\tau)}}, \nonumber \\    
e^{2\,g(\tau)}&=&\tau\coth 2\tau -{\tau^2\over \sinh^2 \, 2\tau}-{1\over   
  4}, \nonumber \\    
\eea    
where,     
\bea    
e_1&=&d\theta_1, \qquad e_2=\sin\te_1 d\p_1, \nonumber \\    
e_3&=&\cos\psi\, d\te_2+\sin\psi\sin\te_2\, d\p_2 -a(\tau) d\te_1,    
\nonumber \\    
e_4&=&-\sin\psi\, d\te_2+\cos\psi\sin\te_2\, d\p_2 -a(\tau) \sin \te_1d\p_1,    
\nonumber \\    
e_5&=&d\psi +\cos\te_2\, d\p_2 -\cos\te_1d\p_1, \quad    
a(\tau)={\tau^2\over \sinh^2\tau}.    
\eea     
where $\mu=0,1,2,3$, we set the integration constant  $e^{\phi_{D_0}}= \sqrt{g_s N}$   The 3-form can be obtained as   
\bea  
H^{RR}& =& g_sN \left[ - {1\over 4} (w^1 -A^1)\wedge (w^2 - A^2) \wedge ( w^3-A^3)  + { 1 \over 4}  
\sum_a F^a \wedge (w^a -A^a) \right] \nonumber \\  
\label{a}  
A &=& { 1 \over 2} \left[ \sigma^1 a(\rho) d \t_1  
+ \sigma^2 a(\rho) \sin\t_1 d\phi_1 +  
\sigma^3 \cos\t_1 d \phi_1 \right]  
\eea  
and the one-forms $w^a$ are given by:  
\be  
w^1 + i w^2   = e^{ - i \psi } ( d \t_2 + i \sin \t_2 d \phi_2)  ~,~~~~~~~~~~  
w^3 = d \psi +\cos \t_2  d\phi_2  
\ee  
Note that we use notation where $x^0,x^i$ have dimension of  
length whereas $\rho$ and the angles  
$\t_1,\phi_1,\t_2,\phi_2,\psi$ are dimensionless and hence the  
appearance of the $\alpha'$ in front of the transverse part of  
the metric. Moreover, following the notation of \cite{ls}  
a factor of $g_s N$  is multiplying the $\alpha'$  instead of $N$ that  
appears in \cite{mn}.           
\subsection{Quadratic fluctuations in the MN background}  
 
The position referred to as $r_0$ in  section (\ref{stringconfine}) is  
$\tau=0$.   
Therefore, we will expand the metric around that value. Let us first  
identify some structures in the metric that are similar to the deformed  
conifold considered in the previous section. Notice that $e_1^2+e_2^2$  
is precisely an $S^2$. Moreover, near $\tau=0$ we have that     
$e^{2g}\approx \tau^2+{\cal O}(\tau^4)$. Thus  
$(\tau,e_1,e_2)$ span an $\mathbb{R}^3$ which we parametrize as  
$(\tau_1,\tau_2,\tau_3)$.   
\begin{equation}  
d\tau^2+e^{2g(\tau)}(e_1^2+e_2^2)\longrightarrow d\tau_1^2+d\tau_2^2+d\tau_3^2.  
\end{equation}    
Certainly  $e_3^2+e_4^2+e_5^2$ parametrizes a space that is topologically  
a three sphere fibered over the $S^2$ spanned by $(e_1,e_2)$. However,  
near $\tau=0$ we have a situation very similar to the structure of the  
metric in the deformed conifold. Namely, at $\tau=0$ there we have that:  
$e_5\to g_5,\,e_3\to \sqrt{2} g_4, \,e_4\to \sqrt{2}g_3$ (up to a  
trivial identification $\theta_1\to -\theta_1,\, \phi_1\to -\phi_1$). This allows us  
to identify this combination as a round $S^3$ of radius  
$2$. Subsequently expanding around the classical values of this round  
$S^3$ we obtain::  
\begin{equation}  
e_5^2+e_3^2 +e_4^2 \longrightarrow dy_1^2+dy_2^2+dy_3^2.  
\end{equation}  
In complete analogy with the KS model the $\tau$ directions receive a  
mass term from the classical classical solution     \begin{equation}  
\g^{\tau\tau} g_{00} \del_\tau X^0 \del_\tau X^0 + \g^{\a\b}\del_\a  
X^i\del_\b X^i g_{ii}  
=\left({8e^{\phi_0}\over 9}\a'g_s   N\,\, e^2  
\cos^2\s\right)\,\, \tau^i\tau_i  
\ee  
 
The B-field in this background is zero. We thus have obtained the same  
bosonic spectrum for the quadratic fluctuations around the Regge  
trajectories in both the KS and MN backgrounds.      
\subsubsection*{Fermionic sector}   
 
The treatment of the fermionic sector is very similar to the  
situation with the KS metric. Indeed, at the classical level, without  
including any fluctuation the RR 3-form flux is all directed along the  
$S^3$ directions and equals\footnote{To bring the RR 3-from field  
strength to this form we use the fact that near the origin  
($\tau=0$) the field $A$ in (\ref{a}) is a pure gauge \cite{mn,gpss}.}  
\be  
H^{RR}=-{1\over 4} g_s N w^1\wedge w^2  \wedge w^3.  
\ee  
This implies that that the effective action for the fermion is  
precisely of the same type as in the KS case except that the value of  
$\ell $ is different in  this case and depends on the specific  
parameters of the MN solution.         
\section{Quantum corrected Regge trajectories for spinning strings\\  
in  confining backgrounds}  
\label{qcrt}    
 
The quantization can be done together for both backgrounds since   
their bosonic and fermionic content is identical. It will be useful to trade  
 the parameters of the theories $\varepsilon$ (KS) and $\phi_0, g_s N$ (MN) by 
  the gauge theoretic quantities: String tension, KK masses, etc.   
\subsection{Bosonic corrections}  
 
As can be inferred from the analysis of the expansion in the KS and MN  
backgrounds around the spinning string the main new modification consist  
on the appearance of a massive term for the radial direction $\tau$. The  
equation of motion for the radial fluctuations is 
\be  
\eqlabel{m0}  
\big[\del_\tau^2-\del_\s^2 +2 m_0^2e^2 \cos^2( \s)\big]\d \tau_i=0, 
\ee  
where $i=1,2,3$ and $m_0^2={4\over 9}\,\, {1 \over g_s N \a'} $ for the MN  
solution and  
$m_0^2= {3^{1/3} a_1\over a_0^{2}}\,\,{\varepsilon^{4/3}\over g_sM\a'}  
\,\,{1 \over g_s M \a'}$  
for the KS solution.   
Assuming that the fluctuations take the standard form  
$\d\tau_i= e^{in\tau}\d \tau_i(\s)$ we obtain: 
\be  
\big[{\del^2\over \del \s^2}+(n^2-{2m_0^2e^2\over 
  2})-{m_0^2}\,\,e^2\,\,\, \cos 2 \s\big] \d \tau_i(\s)=0.  
\ee  
The solution to this equations are the Mathieu functions. One of them is  
$C(a,q,z)$ which is even  and the other $S(a,q,z)$ which is odd. It is  
known that there is no simple analytical presentation for the Mathieu  
functions. However, they are very well studied numerically and series  
expansions near $q=0$ are known (see appendix \ref{mathieu}).  Note  
that for $q=0$ the Mathieu functions are simply $\cos \sqrt{a} u$  
and $\sin \sqrt{a} u$. For nonzero $q$ the Mathieu functions are  
periodic only for certain values of $a$.    
To extract some useful physical information we will content ourselves  
with analyzing the effect of the mass term in these  
solutions. Basically, taking $m_0=0$ in (\ref{m0}) reduces the problem  
to flat space. Thus for  practical reasons we will consider how the  
small but nonzero $m_0$ modifies the flat space result. In principle  
we could go up to higher and higher order in $m_0$ since we know the  
values of the eigenfunction and eigenvalues of the Mathieu equation in  
a series expansion in $q=m_0^2/4$ (see appendix  
\ref{mathieu}). The eigenvalues of the operator (\ref{m0}) are   
\be  
\eqlabel{eigenvalues}  
\lambda_{r,n}={n^2}+{m_0^2e^2} + {r^2} + {1\over  
2(r^2-1)}\,{m_0^4e^4 \over 4}+ {\cal O}(m_0^8),  
\ee  
where $r$ and $n$ are integers. The quantity we are interested is  
\be  
\sum\limits_{r,n\in \mathbb{Z}}\lambda_{r,n} \approx  \sum\limits_{r,n\in  
\mathbb{Z}} r^2+n^2 +m_0^2 e^2  = \ldots  
\ee  
The sum of this type of eigenvalues is similar to the situation for a  
massive scalar field yielding the following expression for the zero  
point energy \cite{g0}:  
\be  
\eqlabel{g0det}  
\begin{split}  
\g_0(\m)&={\m\over 2}+\sum\limits_{n=1}^\infty \sqrt{n^2+\m^2}\\  
&={\m\over 2}+\bigg[-{1\over 12}+{\m\over 2}-{\m^2\over  
2}\ln(4\pi\,e^{-\g}) +\sum\limits_{n=2}^\infty (-1)^n{\Gamma(n-{1\over  
2})\over n!\Gamma(-{1\over 2})}\xi(2n-1)\, \m^{2n}\bigg].  
\end{split}  
\ee  
In our particular case $\m=m_0 e$ and we should trust the above  
expression up to the first nontrivial term in $m_0$. 
 
Let us now investigate the implications of these bosonic modes on   the  
Regge trajectories.  Altogether there are five massless bosonic modes and 
three ``massive'' ones. Substituting their contributions into (\ref{vircobi}) 
we get 
\be 
e(E-\bar E)=\frac{\pi}{2}\left(-\frac{8}{12} + 
 3 m_0 e   \right) 
\ee 
By substituting in the previous equation that 
$\bar E= e/(\alpha'), J= e^2/(2\alpha')$, 
we find that the bosonic fluctuations lead to a {\it nonlinear} Regge  
trajectory  
\be 
J \approx \frac{1}{2}\a'(E- {3\pi\over 2} {m_0})^2  + \Delta_b 
= \frac{1}{2}\a'E^2+\a_0 
- \frac {3\pi}2 \a' E m_0 +\Delta_b 
\ee 
Several remarks are in order: 
(i) The term $\Delta_b$, which would have been the quantum correction had  
the bosonic modes been massless,  will be canceled out by an equal term  
from the fermionic fluctuations. 
(ii) The intercept is given by  $\a'_0= \frac{\a'}{2}({3\pi\over 2}m_0)^2$. Recall 
that $m_0$ is proportional  
to the mass of the glueball of vanishing $J$. This is in agreement with  
the form of the Regge 
trajectory where one can interpret the intercept as    
$\a'_0= \frac{1}{2}\a'm^2_{J=0} $. 
(iii) The deviation between our ``Regge'' trajectory and the usual one is the  
$-\frac{3\pi}{2}\a' E m_0$ term.  
 
We have not been completely rigorous in the treatment of the $\delta\tau_i$  
fluctuations. The complete contribution of these modes to the path  
integral can be computed as in section  
\ref{cov}. Namely, consider the path integral 
\be   
\eqlabel{dresseddet}  
\begin{split}  
&\int D\d\tau_i\,\exp\bigg[-\frac {T_s}2\int d\s d\tau \sqrt{\g} \g^{\a\b}   
\left(\partial_\a \delta \tau_i \partial_\b \delta\tau^i   
+m_0^2 \cos^2\s\d\tau_i\d\tau^i\right)\bigg]\\  
&=\det\left(\Delta_\g-{1\over \sqrt{\g}}m_0^2 e^2\cos^2\s\right)^{-3/2},  
\end{split}  
\ee  
where the determinant is as usual de 2-d Laplace operator:   
\be  
\Delta_\g= {1\over \sqrt{\g}}  \partial_\a \g^{\a\b} \sqrt{\g}\partial_\b.  
\ee  
In the particular conformal gauge where we work, with the worldsheet metric  
equal to the induced target space metric, we can relate the determinant  
of the above expression (\ref{dresseddet}) to the determinant  
computed in (\ref{g0det}). Explicitly we have  
\be  
\det\left(\Delta_\g-{1\over \sqrt{\g}}m_0^2 \cos^2\s\right) 
=\det\bigg[\g^{-1/2}\left(\eta^{\a\b}\pd_\a\pd_\b -m_0^2 e^2 
  \cos^2\s\right)\bigg].  
\ee  
The contribution of the conformal factor was calculated in section  
\ref{cov} following the methods of \cite{dgt,conformaldet} and amounts to a  
shift in the expression of the determinant by -$T/6$ (see equations  
(\ref{fermion}) and (\ref{boson})).

\subsection{Fermionic corrections}   
 
To complete our treatment of Regge trajectories  
in confining backgrounds we now turn to fermions.   
We will fix, as in the flat space case, the  
kappa-symmetry by identifying $\theta^1=\theta^2=\theta$.  
The fermionic action, to terms 
quadratic in the fluctuations, is given by 
\be  
S_F\approx \frac {i}{2} T_s \int\sqrt\g \g^{\a\b} \left( 
\bar{\theta}\partial_\a \bar X^\mu \Gamma_\mu \partial_\beta\theta  
+\frac 14\partial_\a \bar X^\mu\partial_\b \bar X^\nu \bar{\theta}  
\gamma_\mu \hat f 
\gamma_\nu\right)  
\ee  
where $\hat f$ is the contraction of the 3-form field strength with  
ten dimensional Dirac gamma matrices. 
In particular, we have 
\be 
\hat f^2 = -\frac32 \frac{1}{g_s^3 M\a' \, a_0^{3/2}}\label{hatf}. 
\ee 
 
In the conformal gauge the action becomes 
\bea 
S_F&\approx& \frac i2  
T_s \int \left(\eta^{\a\b}\partial_\a \bar X^\mu \bar\theta 
\Gamma_\mu \partial_\beta \theta - \frac 12 \sqrt\g \bar\theta\hat f\theta 
\right) 
\nonumber\\ 
&=&\frac i2 T_s\int \bigg[e \sqrt{ g_{00}}\bigg(\bar\theta 
(\Gamma^0 - (\Gamma^1\cos\tau-\Gamma^2\sin\tau)\sin\s)\dot\theta 
\nonumber\\ 
&+&\bar\theta( \Gamma^1\sin\tau+\Gamma^2\cos\tau)\cos\s \theta' \bigg) 
-\frac 12 \sqrt\gamma \bar \theta \hat f \theta\bigg]. 
\eea  
The kinetic term can be simplified further by using the same type of  
unitary transformation as in the flat  
space case (see (\ref{rotation})), at the expense of obtaining 
a more complicated mass term. In the end, one obtains 
\be 
S_F\approx \frac i2 
T_s\int \cos\s \bigg[e \sqrt{ g_{00}} \bigg(\bar\Psi \Gamma^0\dot \Psi+ 
\bar\Psi \Gamma^2 \Psi'\bigg) + \Psi\theta \calm \Psi\bigg], 
\ee 
where the mass term is  
\be 
\calm=-\frac 1{2\cos\s} \sqrt\gamma \hat f + e\sqrt{g_{00}}\bigg( 
\Gamma^0 U_{01}^{-1} U_{12}^{-1} \; \partial_\tau U_{12} U_{01} +  
\Gamma^2  U_{01}^{-1}\; \partial_\s U_{01}\bigg). 
\ee 
The ten dimensional rotations implemented by 
\be 
U_{12} = e^{\frac 12\tau \Gamma_1\Gamma_2}\qquad U_{01} =  
\exp(-\frac 12 {\rm arccosh}(\frac{1}{\cos\sigma}) \Gamma_0\Gamma_1) 
\ee 
are used to rewrite the Dirac matrices that appear in the kinetic term as 
\bea 
&&\Gamma_1 \cos\tau-\Gamma_2\sin\tau=U_{12}U_{01} 
{\Gamma_1}U_{01}^{-1}U^{-1}_{12} \nonumber\\ 
&&U_{12}U_{01} \Gamma_0\cos\sigma U^{-1}_{01}U^{-1}_{12} 
= (\Gamma_0 + (\Gamma_1\cos\tau-\Gamma_2\sin\tau)\sin\s 
=U_{12}U_{01} \Gamma_0\cos\sigma U^{-1}_{01}U^{-1}_{12}.  
\eea 
 
\subsubsection*{The fermionic determinant}   
 
More rigorously and to avoid issues with the field redefinition of the  
fermions as in (\ref{redef}) we turn to the path integral evaluation of  
the partition function. The path integral we want to consider is similar to the one for  
fermions in flat space with the notable difference that the RR 3-form  
field strength is nonvanishing and contributes what we could think of as   
an extra  mass term for the fermions  
\bea  
&&\int [D\theta ] \;\exp\bigg[\frac {iT_s}2 
\int d\sigma d\tau \sqrt{\g}   
\bar\theta \g^{\a\b}\partial_\a \bar X^m \Gamma_m  
\left(\partial_\b+{1\over 2}\partial_\b\bar X^n\hat{f}\Gamma_n\right)  
\theta\bigg] \nonumber \\  
&=& \bigg[\det(\g^{\a\b} \partial_\a \bar X^m \Gamma_m \left( 
  \partial_\b +{1\over 2}\partial_\b\bar X^n\hat{f}\Gamma_n\right)\bigg]^8  
\nonumber\\  
&=&\bigg[ \det(\g^{\a\b} \partial_\a \bar X^m \Gamma_m \left(  
\partial_\b +{1\over 2}\partial_\b\bar X^n\hat{f}\Gamma_n\right)  
\g^{\g\d} \partial_\g \bar X^q \Gamma_q \left( 
  \partial_\d +{1\over 2}\partial_\d\bar 
  X^p\hat{f}\Gamma_p\right)\bigg]^4 \nonumber \\  
&=& \bigg[\det\left(\g^{\a\b} \partial_\a\partial_\b + 
  \hat f^2 \right)\bigg]^4 \nonumber \\  
&=& \bigg[\det\,\, \g^{-1/2}\left(\eta^{\a\b} \partial_\a\partial_\b - 
  m_F^2\right)\bigg]^4.  
\eea  
where the fermionic mass term $m_F$ can be expressed as  
\be 
m_F^2 = \frac{3 \varepsilon^{4/3}}{2^{4/3} a_0^2 (g_s^4 M^2\a'^2)} e^2 \cos^2\s 
\equiv 2 l^2 e^2 \cos^2\s 
\ee 
 
\subsection{Quantum corrected Regge trajectories from string theory} 
 
Putting together all the partial results of this section we obtain that  
the corrections to the classical Regge trajectories is given by the zero point  
energy:   
\be  
\eqlabel{theformula}  
\ln Z = \pi\left(4 \ell-\frac {3 }2 m_0\right). 
\ee   
This explicit formula has been obtained in the limit in which both  
$m_0$ and $\ell$ are small. This is the limit in which the  
eigenvalues  of the Mathieu equation have explicit analytical  
expressions. In a more general situation there is no reason for us to  
restrict ourselves to the small $m_0$ and $\ell$ limit. In fact, as  
mentioned before, general formulas for the eigenvalues of the Mathieu  
equations are available.     
\begin{table}[hbt]  
\begin{center}      
\begin{tabular}[h]{|l||c| |c|}      
\hline     
 {}& Klebanov-Strassler  & Maldacena-N\'u\~nez \\     
\hline       
\hline       
$m_0 $&$ {3^{1/6}a_1^{1/2}\over a_0}\,\,\,\,  
\frac{\ep^{2/3}}{{g_s M\a'}} $ & 
 $ {2 \over 3}\,{1 \over \sqrt{g_s N \a'}}$ \\      
\hline       
$\ell $&$ {3^{1/2}\over 2^{7/6}\,a_0}\,\,\,g_s^{-1}  
{\varepsilon^{2/3} \over {g_s M\a'}} 
 $ & $ {2^{1/2} \over g_s N  \sqrt{g_sN\a'}} $\\      
\hline       
\hline       
\end{tabular}      
\caption{Parameters determining the Regge intercept for the KS and MN solution.\label{ml}}  
\end{center}      
\end{table}  
According to the expression (\ref{zero}) the effect of these quantum 
corrections on the Regge trajectories is  
\be 
\eqlabel{correction} 
e(E-\bar E)= \left(\frac {3 }2 m_0-4\ell\right) \pi \, e \equiv  z_0 e 
\ee 
 
Then, by substituting in the previous equation that 
$\bar E= e/(\alpha'), J= e^2/(2\alpha')$, 
we derive a {\it nonlinear} Regge trajectory 
\be 
\eqlabel{nt} 
J={1\over 2} \a'_{eff} \,\,\,E^2 - \a'_{eff} z_0 \,\, E +{1\over 2} 
  \a'{}_{eff}z_0^2.  
\ee 
Based on table (\ref{ml}) and the expression (\ref{theformula}), it  
seems natural  to consider the small $\ell$ limit since  for both 
backgrounds the dimensionless effective parameter is $\ell\,e$. This 
combination is proportional to the classical string energy  and 
therefore controls the backreaction of the spinning string on the  
supergravity background. In the case of the KS 
background there is another ratio: the deformation parameter to the 
effective radius of the background, which can be taken to be 
small. The supergravity limit requires the effective radius of the 
background in string units $g_s N$ to be large. There are factors of $g_s$ in both  
expressions which can in principle be taken arbitrary and thus allows  
us to explore other regimes but this might require considering higher loops.   
Nevertheless, the case can be made for  
generic values of $\ell$ and $m_0$ resulting in a {\it positive value} of  
the expression (\ref{theformula}). This implies that the second term 
in (\ref{nt}) is negative and that the intercept of the corresponding 
Regge trajectory is {\it positive.}

\section{Phenomenology}  
\label{pheno}   In this section we attempt to elucidate how our results compare to the  
phenomenological data. Let us begin by clarifying the regime of validity  
of our string theory calculations. The models we considered are expected  
to be dual to ${\cal N}=1$ SYM plus extra matter coming from the KK  
supergravity modes. Evidently this is not QCD, which will constitutes  
our main source of experimental and lattice results. In the  
semiclassical   
analysis that we use, we assume that  
the supergravity approximation is valid. The validity of SUGRA has two  
implications. The first one is that the extra   
matter fields and the pure glue fields will have approximately similar  
masses and we will necessarily end up with mixing of these two  
sectors. The second implication is that the validity of the  
supergravity approximation requires low curvatures which translates into  
large $N$ in the gauge theory.  We will  
nevertheless, see remarkable qualitative and even quantitative  
similarities.   Our last cautionary point involves the difference among Regge  
trajectories for mesons, baryons and glueballs. For mesons and baryons    
Regge trajectories have been experimentally  
confirmed since the 60's.   
The precise experimental value of the slope $\a'$ depends on the flavor  
content of the states lying on the corresponding trajectory. However,  
a universal value could be taken to be $\a'\approx .85 GeV^{-2}$. For  
the purpose of this paper we will consider some of the trajectories   
presented in \cite{collins}  but will use the current (2002) particle   
data book for accuracy. The soft Pomeron trajectory is qualitatively different from the Regge  
trajectories for mesons and baryons. Its slope is flatter. The  
phenomenological parameters of the soft Pomeron are: $J=1.08+ 0.25 t$,  
that is $\alpha'=0.25$ GeV$^{-2}$ \cite{pomeron}. The identification  
of the Pomeron with glueballs seems very plausible  
and a strong pushed toward its demonstration is being made using lattice  
\cite{gluepomeron} and other semi-analytical techniques  
\cite{gluepomeron1}. The identification of the soft Pomeron trajectory 
with a trajectory of glueballs does not provide direct information 
about the value of the glueballs and we therefore turn to the more 
complete data provided by lattice results  \cite{gluelat1,gluelat2}.        
There are many models attempting to explain the difference between the  
Pomeron slope and the slope for baryons and mesons. Most of the  
arguments are based on the universal idea that the flux tube between a quark  
and an antiquark has flux in the fundamental representation while for glueballs the  
flux is in the adjoint representation. It then follows that the ratio of the slopes is  
given by the Casimir operators in the fundamental and adjoint representations:  
\be  
{\alpha_{gg}\over \alpha_{q\bar q}}={C_{fund.}\over 
  C_{adjoint}}={N^2-1\over 2N^2}  
\ee   In the large N limit the ratio goes to 1/2. This is precisely the  
ratio of slopes obtained by considering the trajectories resulting  
from classical solutions of closed and open strings spinning in flat  
space  (see eqn (\ref{rego}) and (\ref{regc})). Since, as shown in section \ref{classical},  
at the classical level   
the net  effect of putting the strings to spin in actual confining  
backgrounds rather than in flat space, is a rescaling of the string   
tension we conclude that this  
1/2 is precisely the ratio of the slopes of the Regge trajectories for  
closed/open  (glueball/mesons) predicted by the gauge/gravity correspondence.      
\subsection{A theoretical value of the Regge slope for glueballs}  
In this subsection we will show that the low-lying glueball  masses calculated  
in the KS model by C\'aceres and Hern\'andez  
\cite{ksglue} provide an impressive  numerical match for the slope of  
the soft Pomeron trajectory. A  similar analysis was carried out 
for the MN background in \cite{pere}, it results turn out to be less 
conclusive. The values obtained  in \cite{ksglue} are presented in  
table \ref{ksglueballs}.    \begin{table}[hbt]  
\begin{center}      
\begin{tabular}[h]{|l||c|}      
\hline     
 State& (Mass)$^2/\varepsilon^{4/3}$ \\     
\hline       
\hline       
$0^{++}$ & 9.78 \\      
\hline       
$0^{++ *}$ & 33.17 \\      
\hline       
$1^{--}$  &  14.05\\      
\hline       
$1^{-- *}$  & 42.90 \\      
\hline       
\end{tabular}      
\caption{Mass$^2$ in units of the conifold deformation   
for the low-lying glueballs in the KS model \cite{ksglue}.  
\label{ksglueballs}}  
\end{center}      
\end{table}   The mass is measured in units of the conifold deformation  
$\varepsilon^{2/3}$. In the KS background the conifold deformation  
naturally sets the mass scale of the four dimensional gauge theory.    
However, there is also a  
large number $g_s M$ which sets a hierarchy of scales scales between the  
glueball masses and the string tension \cite{ks}. Since we do not reliably know how to fix the value  
of $\varepsilon$ it makes little sense to find the full Regge trajectory  
$J=\alpha_0+\alpha' t$. However, since the combination $\alpha'\, t$ is  
independent of the units used to measure mass, it makes sense to  
compute it. Using, as customary,    
the lightest glueballs for a given spin we find  
\be  
J=0.234\, t + \alpha_0.  
\ee  
The first term is remarkably close to the experimental value for  
the soft Pomeron of $0.25\,t$.

\subsection{Glueball Regge trajectories from the lattice}  
Since definite experimental evidence for the existence of glueballs  
remain elusive we will limit our analysis to the lattice data.   
The lattice results we will use are extracted from  
\cite{gluelat1,gluelat2}. The analysis of \cite{gluelat1} was  
performed for QCD $(N=3)$, the prediction for the lowest lying  
glueball at each spin are given in table \ref{MP}   \begin{table}[hbt]  
\begin{center}      
\begin{tabular}[h]{|l||c|}      
\hline     
 State& Mass (GeV)  \\     
\hline       
\hline       
$0^{++}$ & 1.73 \\      
\hline       
$2^{++ }$ & 2.40 \\      
\hline       
$3^{++}$  & 3.64\\      
\hline       
\end{tabular}      
\caption{Continuum limit for glueball masses from Morningstar and  
Peardon \cite{gluelat1}.\label{MP}}  
\end{center}      
\end{table}   The analysis of \cite{gluelat2} is summarized in table \ref{T}. It  
contain extrapolations of the values of the mass for $N\to \infty$  
assuming that the difference between the masses at finite and infinite  
$N$ is of the form ${\rm const.} \,N^{-2}$.     \begin{table}[hbt]  
\begin{center}      
\begin{tabular}[h]{|l||c| |c|}      
\hline     
 State& Mass(GeV) for $SU(3)$ & Mass(GeV) for $SU(N\to \infty)$ \\     
\hline       
\hline       
$0^{++}$ & 1.64 & 1.60\\      
\hline       
$2^{++ }$ & 2.33 &2.16. \\      
\hline       
$3^{++}$  & 4.00 & ---\\      
\hline       
\end{tabular}      
\caption{Glueball masses from Teper \cite{gluelat2}.\label{T}}  
\end{center}      
\end{table}   The best linear fit for the data of \cite{gluelat1} and  
\cite{gluelat2} is given by    \be   
J_{MP}=-0.234+0.259\,\, t_{MP}, \qquad \quad J_T=0.148+.189\,\,t_{T}.  
\ee  
As we have seen in previous sections the intercept is a crucial  
parameter that result from quantum corrections on the string side.   
We are thus particularly interested in comparing our theoretical  
result against the lattice date. Unfortunately, the evidence is  
inconclusive since the data of \cite{gluelat1} yields a negative  
intercept while \cite{gluelat2} yields a positive intercept.   
If we assume that all the points have roughly the same error we could  
combine all the data, this results in   
\be  
J= 0.0265 + 0.213\,\, t.  
\ee  
This trajectory has a positive slope and the value of $\alpha'$ is  
closer to the one observed for the soft nonperturbative  
Pomeron. Another argument in favor of a positive intercept can be  
given by considering the best fit without including the zero spin  
states. The reason for excluding the zero spin states arises because  
the semiclassical analysis requires large spin.

\subsection{Pomeron phenomenology }  
\subsubsection{The intercept} 
There are other considerations leading to a positive intercept for the  
soft Pomeron trajectory.   
A given Regge trajectory $\a(t)$ contributes a term proportional to  
$s^{\a(0)-1}$ to the total cross section $\sigma^{ab}$.   
It is observed that many  total cross sections in elastic scattering $(\bar{p}p, \, pp,\,  
\pi^- p, \pi^+ p)$ rise with $s$ at high  
energy \cite{pomeron}, which is why a soft-pomeron-exchange term is needed in  
addition to the experimentally  observed trajectories of  
$(\rho,\omega,\ldots)$.   
One of the implications of Regge theory is that the total  
cross-section has, to leading order in $(s/s_0)$,  a simple power   
behavior: $\sigma^{tot} \sim  
\left({s\over s_0}\right)^{\a(0)-1}$, where $s_0$ is empirically  
around 1 Gev$^2$.  The total cross-sections for various states are remarkably constant  
over a large range of $s$. Constancy of $\sigma^{tot}(s)$ requires  
$\a_0\approx 1$. The fact that most mesonic and baryonic trajectories  
have $\a_0$ different from 1 prompted introduction of a new trajectory  
with $\a_0\approx 1$.   
In \cite{pomeron} a parametrization of the intercept for the Pomeron  
trajectory as $\epsilon=\alpha(0)-1$ has been shown to fit all data  
for total cross sections $(\bar{p}p, \, pp,\,  
\pi^- p, \pi^+ p,\, K^- p, \, K^+p,\, \bar{p}n, \, pn)$ with $\epsilon  
= 0.081$ and $\epsilon  
= 0.096$  
This value of the intercept which imply a positive power of $s$ in the  
total cross section will naively violate the Froissart bound  
$\sigma^{ab} \le (\pi/m_{\pi}^2) \log^2 (s/s_0)$ when the energy  
becomes extremely large. It is natural to assume then, that the power  
of $s$ is only an effective power which must reduce as $s$  
increases. There are various proposal as to the concrete mechanism for  
achieving this variation in the power of $s$ including different  
couplings of the pomeron and or multipomeron exchange \cite{pomeron,pomeron1}.   
However, the explanation is that  
the total cross section takes this form only in the range of energies  
discussed, for higher energies there should be two or more pomerons  
being exchanged.    It is, of course, far-fetched for us to claim that our formula  
(\ref{theformula}) coincides with the experimental data. What is natural  
to assume, given that the metric and the three form are related by the  
Sugra equations of motion, is that $m_0$ and $\ell$ in  
(\ref{theformula}) are of the same order. Under this natural assumption  
we obtain {\it a positive intercept} for the Regge trajectories  
describing glueballs.     
 
\subsubsection{Nonlinearity of the Pomeron Regge trajectory from string theory}  
There are various experimental and theoretical reasons why the Regge 
trajectory must be nonlinear.  Strong  experimental evidence for the nonlinearity of the 
 Pomeron Regge trajectory was presented by the UA8 collaboration \cite{nlex}. The standard linear trajectory 
of the Pomeron is too small in the 1-2 GeV${}^{-2}$ $|t|$ - region to 
adequately describe the data and it is suggested experimentally 
that: 
\be 
\eqlabel{pomenl} 
\alpha(t) = 1.10 + 0.25 t + \a''\,\, t^2,  
\ee 
where $\a'' = 0.079 \pm 0.012 GeV^{-4}$. It is, of course a challenge 
to provide a theoretical foundation for such trajectory. Some 
phenomenological attempts have been put forward in, for example, 
\cite{nl2}. We find it very encouraging that the nonlinear 
trajectories we obtain share some of the properties of (\ref{pomenl}) 
at the qualitative level. In particular (\ref{nt}) has a positive 
intercept and positive curvature ($\alpha(t)''>0$) which are  completely compatible with 
(\ref{pomenl}).

\subsection{Regge trajectories for mesons}  
The data for mesons is more abundant. Some of the trajectories are  
\begin{itemize}    
\item $\rho(770)\,(1^{--}),\quad  a_2(1320)\,(2^{++}),\quad  
\rho_3(1690)\,(3^{--}),\quad a_4(2040)\,(4^{++})$   
\item  $\o(782)\,(1^{--}),\quad f_2(1420)\,(2^{++}),\quad  
\o_3(1670)\, (3^{--}),\quad f_4(2050)\,(4^{++})$    
\item $K^*(892)\, (1^-),\quad K^*_2(1430)\,(2^+),\quad  
K^*_3(1780)\,(3^-),$   \quad $K^*_4(2045)\, (4^+),\quad  
K^*_5(2380)\,(5^-)$    
\item $\pi^0(135)\, (0^{-+}),\quad b_1(1235)\, (1^{+-}),\quad  
a_2(1700)\, (2^{++})$.      
\end{itemize}  
In the above list the masses are given in MeV but the Regge  
trajectories are usually written in GeV units.   
The main conclusion we would like to draw is that the  
intercept depends on the particular Regge trajectory but it is  
generically small and positive. It can also be seen that the slope is  
practically universal and considerably larger than that of the  
glueballs.  Our main analysis  applies strictly to Regge trajectories made of  
glueballs since we consider closed strings. However, the treatment of  
open strings is similar. The technical elements follow closely those  
described in this paper which yield a positive value for the 
intercept.    
 
\subsubsection{Nonlinear Regge trajectories for mesons} 
Realistic Regge trajectories extracted from data are {\it 
nonlinear}. For example, the straight line that crosses the $\rho$ 
and $\rho_3$  squared masses corresponds to an intercept of 
$\a_\rho(0)=0.48$, whereas the physical intercept is located at 
$0.55$. Similarly,  the straight line which crosses the $K_2^*$ and $K_4^*$ 
squared masses corresponds to an intercept of 0.1, whereas the physical 
intercept is 0.4.  A more detailed studied was carried out in  \cite{nl1}. 
 
The corrections computed using strings spinning in sugra backgrounds 
dual to confining gauge theories have a similar qualitative effect  on 
linear Regge trajectories.

\subsection*{Comments on the 2+1 Glueball trajectories}  
It is interesting to note that the existence of Regge trajectories for  
glueballs in 2+1 dimensions has been shown  explicitly in  
\cite{3dglue}. In this case, as opposed to the cases considered in 3+1,  
the intercept is unambiguously negative.  It would be very  
interesting to extend the analysis of this paper to 2+1 where lattice data is  
more abundant.  
In the context of the gauge/gravity correspondence there  
are supergravity backgrounds dual to confining 2+1 gauge theories. It  
has recently been shown that these  Sugra backgrounds dual to confining  
gauge theories in 2+1 admit a set of hadronic states similar to the  
annulons \cite{italia}. This fact encourages one to believe that more  
generic hadronic states similar to the large spin hadrons considered  
in this paper  ought to exists in these 2+1 confining theories.              
 
\section{Conclusions}  
In this paper we have studied Regge trajectories from the string 
theory point of view. Classically, Regge trajectories in string theory 
are linear with zero intercept. We have shown that for strings 
spinning in flat space the one-loop corrected Regge trajectory 
receives an positive intercept in the bosonic case whereas in the 
supersymmetric case the trajectory remains a straight line through the 
origin even at one loop order.

Our main results are related to the analysis  of strings spinning in 
supergravity backgrounds dual to confining gauge theories. By 
explicitly considering the KS and MN backgrounds we obtained that for 
configurations of spinning string they are both qualitatively 
identical. A unified analysis of the quantum correction was carried 
out in section \ref{qcrt}. Generically the trajectories are nonlinear 
at the one loop level. We found that for the full background the 
resulting Regge trajectory generically has a positive intercept and 
the has $\alpha(t)'' >0$. Since we consider spinning closed strings, 
our results should be compared to glueball trajectories. The most 
relevant experimental information comes from the soft Pomeron 
trajectory. The Pomeron trajectory has a positive 
intercept. However we find that the modifications to the Regge 
trajectories are different for the bosonic string and the 
superstring. Motivated by an attempt to see if our analysis contains 
in principle a qualitative match we considered only the bosonic 
contribution at one loop. Interestingly we find that two distinctive 
features of our trajectory coincide with properties of the he  
best experimental fit to the UA8 
collaboration data \cite{nlex}, that is a Regge trajectory of the form: 
$J=\alpha(t)=\alpha(0) + \alpha' \,\, t + \alpha'' \,\,t^2$. Our 
trajectories have a positive intercept and a positive curvature 
$(\alpha(t)''>0)$.

We would like to comment of the regime of validity of our 
calculations. Since we use a classical string in supergravity  backgrounds we 
start by requiring that the supergravity approximation to string theory  
be valid. That implies small curvatures which in the field theory side 
means large rank of the gauge group $N$. The fact that we work at 
large $N$ implies that some finer structure of Regge trajectories 
(cuts, etc.) is hidden. Making contact with this finer structure is 
one of the most interesting directions of future development. Equally 
worth pursuing  is the precise  relation of our work to the 
study of Polchinski and Strassler on  Regge physics 
\cite{pstr} (Regge physics in the context of the AdS/CFT has also been 
discussed in \cite{peschanski}). Another important point in our approximation is that we 
treat the string semiclassically. This implies that we want its energy 
to be large so that the fluctuations are reasonably small but we 
certainly do not want to make it so heavy as to invalidate treating it 
as a probe in the supergravity background. This is easily achieved in 
the models we consider and implies a hierarchy between the glueball 
mass and the curvature of the supergravity background. Let us also 
point out that  the models which we discussed are dual to gauge theories with no 
fundamental matter and this  limits further the comparison to QCD. It 
would very interesting to improve our discussion by considering models 
with fundamental matter.

It is worth mentioning that our result is not only compatible with 
some of the aspects of the best experimental fit to the Pomeron 
trajectory but seems to be compatible with some of the 
phenomenological models introduced in the literature, in  
particular, models that contain square roots.

\section*{Acknowledgments}  
We thank  D. Chung, J. Liu, M. Kruczenski, G. Sterman  and E. Yao for useful 
comments,  
E. Gimon for discussions and participation in  
the initial stages of this project. We thank D. Mateos for correcting a 
statement about supersymmetry in a previous version of the 
manuscript. We are thankful to A. Tseytlin for  
various clarifications and comments. We are particularly thankful to 
I.  Klebanov for illuminating comments. J.S. and D.V. would like  
to thank MCTP for hospitality during the initial  
stages of this work.  LAPZ is partially supported by 
DoE. J.S would like to thank A. Font and S. Theisen for useful discussions. 
The work of J.S was  supported in part by the German-Israel Foundation  
for Scientific Research and by the Israel Science Foundation.  
D.V. is supported by DOE grant DE-FG02-91ER40671.

\appendix  
\section{Evaluation of the free energy for the static open string}  
\label{fe}   We include here the details of the evaluation of the free energy for the  
static open string. Our purpose is to show explicitly how the  
regularization works. We follow \cite{alvarez} who uses analytic  
regularization in which   
\be  
\ln x =-{\del\over \del \b} x^{-\b}|_{\b=0}\,.  
\ee  
We want to evaluate   
\be  
\det \left(-\del_\tau^2 -\del_\s^2\right)= \exp {\rm Tr}\, \ln(-\del_\tau^2 
  -\del_\s^2).  
\ee  
The fluctuations have to vanish at the boundary and therefore the  
eigenfunctions are of the form $\eta(\tau,\sigma)=\sin(n\pi \s /L)\,  
\sin(m\pi \tau /T) $ with eigenvalues    \be  
\l_{m,n}=\left({m\pi\over T}\right)^2 + \left({n\pi\over L}\right)^2.  
\ee   Thus,   
\be  
\begin{split}  
{\rm Tr}\, \ln(-\del_\tau^2 -\del_\s^2)&= \sum\limits_{m,n}\ln \l_{m,n} \\  
&=\sum\limits_{m,m}-{\del\over \del \b}(\l_{m,n})^{-\b}|_{\b=0} \\  
&=-{\del\over \del \b}\sum\limits_{m,n}\left(({m\pi\over T})^2   
+({n\pi\over L})^2\right)^{-\b}|_{\b=0}.  
\end{split}  
\ee  
We now trade the sum over $m$ by an integral, given that we are  
interested in the limit $T\to \infty$   \be  
\begin{split}  
&-{\del\over \del \b}\sum\limits_{n=1}^\infty {T\over 2\pi}\int d\o  
\left(\o^2+(n\pi/L)^2\right)^{-\b} |_{\b=0} \\  
&=-{\del\over \del \b}{T\over 2\pi}\sum\limits_{n=1}^\infty  
\left({n^2\pi^2\over L^2}\right)^{1/2-\b} \sqrt{\pi}   
{\G(\beta-1/2)\over \G(\beta)}|_{\b=0} \\  
&-{T\over (4\pi)^{1/2}}{\del\over \del \b}{\G(\b-1/2)\over 
  \G(\b)}\xi(2\b-1)\left( { L^2\over \pi^2}\right)^{\b-1/2} |_{\b=0} \\  
&-{\pi \,\, T \over 12\,\, L}.  
\end{split}  
\ee           
\section{Some properties of Mathieu functions}  
\label{mathieu}  
In this appendix we collect some known facts about Mathieu functions  
that are relevant for the concrete use made in the main text. The  
Mathieu differential equation is   
\be  
{d^2\over dz^2}y+[a-2q\cos(2\,z)]y=0.  
\ee  
Its general solution is of the form   
\be  
y=c_1 C(a,q,z)+c_2 S(a,q,z),  
\ee   
where $C(a,q,z)$ and $S(a,q,z)$ are Mathieu functions denoting the  
even and odd solutions. As mentioned in the main test, there is no  
analytic simple form for these functions. However series expansions  
for the functions $C(a,q,z)$ and $S(a,q,z)$ near small $q$ are well  
known.   
\be  
\begin{split}  
C(a_r(q),q,z)&\approx \cos(r\,z)+{1\over 4}\left({\cos((r-2)z)\over  
r-1}-{\cos((r+2)z)\over r+1}\right)\,q \\  
&+{1\over 32}\left({\cos((r-4)z)\over (r-2)(r-1)}-{2(r^2+1)\cos( rz)\over  
(r^2-1)^2}+ {\cos((r+4))z \over (r+1)(r+2)}\right)\, q^2 +\ldots  
\end{split}  
\ee  
here $r$ is an integer number known as the characteristic exponent of  
the Mathieu function which allows to write the Mathieu function as  
$\exp(i\, r\, z) f(z)$ where $f(z)$ is $2\pi$ periodic with  
characteristic value $a$.    There are also well established series expansions for the  
Mathieu characteristics, that is, for the values of $a$ such that the  
Mathieu function is  periodic.   
\be  
\begin{split}  
a_r(q)&\approx r^2 + {1\over 2(r^2-1)}\,q^2+ {5r^2+7\over  
32(r^2-4)(r^2-1)^3}\,q^4\\  
&+{9r^4+58r^2+29\over 64(r^2-9)(r^2-4)(r^2-1)^5}\,q^6+ \ldots  
\end{split}  
\ee

\end{document}